\documentclass[a4paper,10pt]{article}
\usepackage{amsmath,mathtools}
\usepackage[width=14.5cm,height=24cm]{geometry}

\usepackage{xcolor}
\usepackage{amsmath,amscd}
\usepackage{amssymb}

\usepackage{graphicx}
\usepackage{caption}
\usepackage{subcaption}

\usepackage{tikz}
\usetikzlibrary{arrows}
\usetikzlibrary{decorations.markings}

\catcode`@=11
\def\frownfill{$\scriptscriptstyle\m@th\mathord\frown\mkern-0.2mu
  \cleaders\hbox{$\mkern-2mu\smash-\mkern-2mu$}\hfill
  \mkern-0.2mu$}
\def\bow#1{\vbox{\m@th\ialign{##\crcr
      \frownfill\crcr\noalign{\kern-0.2\p@\nointerlineskip}
      $\hfil\displaystyle{#1}\hfil$\crcr}}}
\def\bbow#1{\vbox{\m@th\ialign{##\crcr
     \frownfill\crcr\noalign{\kern-0.7\p@\nointerlineskip}
     \frownfill\crcr\noalign{\kern-0.3\p@\nointerlineskip}
      $\hfil\displaystyle{#1}\hfil$\crcr}}}
\def\widefrownfill{$\m@th\mathord\frown$}
\def\widebow#1{\vbox{\m@th\ialign{##\crcr
      \hfil\widefrownfill\hfil\crcr\noalign{\kern-0.9\p@\nointerlineskip}
      $\hfil\displaystyle{#1}\hfil$\crcr}}}
\def\widebbow#1{\vbox{\m@th\ialign{##\crcr
     \hfil\widefrownfill\hfil\crcr\noalign{\kern-1.8\p@\nointerlineskip}
     \hfil\widefrownfill\hfil\crcr\noalign{\kern-0.9\p@\nointerlineskip}
      $\hfil\displaystyle{#1}\hfil$\crcr}}}

\begin{document}
{\centering
{\Large
Discretization of Maxwell's Equations for Non-inertial Observers Using Space-Time Algebra}
\\[2em]

Mariusz~Klimek\textsuperscript{*, 1}, 
Stefan~Kurz\textsuperscript{$\P$, 1},
Sebastian~Sch\"ops\textsuperscript{$\dagger$, 1, 2} 
and Thomas~Weiland\textsuperscript{$\ddagger$, 2}
\\[2em]

\textsuperscript{1} Graduate School of Computational Engineering, Technische Universit\"at Darmstadt, 
Dolivostra\ss e 15, D-64293, Darmstadt, Germany \\
\textsuperscript{2} Institut f\"ur Theorie Elektromagnetischer Felder, Technische Universit\"at Darmstadt, 
Schlo\ss gartenstra\ss e 8, D-64289 Darmstadt, Germany
\\[2em]

* Corresponding author, E-mail: klimek@gsc.tu-darmstadt.de,\\ Phone: +49 6151 16-24391, Fax: +49 6151 16-24404 \\
$\P$ kurz@gsc.tu-darmstadt.de \\
$\dagger$ schoeps@gsc.tu-darmstadt.de \\
$\ddagger$ thomas.weiland@temf.tu-darmstadt.de \\
}
\vspace{5em}

\textbf{Abstract:}
We employ Maxwell's equations formulated in Space-Time Algebra to perform discretization of moving geometries
directly in space-time. All the derivations are carried out without any non-relativistic
assumptions, thus the application area of the scheme is not restricted to low velocities.
The 4D mesh construction is based on a 3D mesh stemming from a conventional 3D mesh generator. 
The movement of the system is encoded in the 4D mesh geometry,
enabling an easy extension of well-known 3D approaches to the
space-time setting. 
As a research example, we study a manifestation of Sagnac's effect in a rotating ring resonator.
In case of constant rotation, the space-time approach enhances the efficiency of the scheme,
as the material matrices are constant for every time step, without abandoning the relativistic framework.

\section{Introduction}

Numerical simulation of electromagnetic field in a rotating medium has been studied 
in frequency-domain \cite[and references therein]{Steinberg},
as well as in time-domain \cite{SagnacFDTD,Novitski}.
In all these approaches it is assumed from the beginning that the angular velocity $\Omega$ is small,
i.e., $\frac{r \Omega}{c} \ll 1$ with $r$ the distance from the centre of rotation,
and in \cite{Steinberg} additionally $\frac{\Omega}{\omega} \ll 1$ with $\omega$ the frequency of the electromagnetic field.
These assumptions entail several drawbacks.
First,
they limit the range of applicability, e.g., 
simulation of circular accelerators with respect to a rotating observer cannot be carried out.
Second,
to estimate the error committed in a low velocity approximation,
one certainly needs a fully relativistic description.
Third, 
the physical interpretation is not clear.
Quantities as time and electromagnetic fields are assumed to be the same in a rotating and a stationary reference frame.
One may try to justify this assumption by the fact that the difference due to relativistic effects is small, thus negligible.
However, the quantities of interest, such as the rotation-induced frequency shift, are also small.
Therefore, one needs to distinguish between quantities in different reference frames in order to interpret
the computed quantities properly.

Since Maxwell's equations are invariant with respect to Lorentz transformations, it calls for the relativistic treatment.
The coordinate and metric free description of relativistic electromagnetism is outlined in \cite{Kurz}.
In this paper we develop a fully relativistic computational scheme for non-inertial observers.
The scheme provides physical underpinning for further simplifications and enables the possibility to account
for the error of such simplifications.
In principle one could apply the scheme proposed in \cite{simplices4D} to simulate the fields in a rotating reference frame.
However, this would require a 4D simplicial mesh generator, would not allow simple, explicit time-marching and
might require more computational resources,
and one would also need to employ the formalism of homology/cohomology groups as
they are not trivial even in the case of a relatively simple structure as a ring resonator considered in this paper.

We employ Clifford's Geometric Algebra (GA)
associated with Minkowski space-time; this is known as Space-Time Algebra (STA).
By imposing Minkowski metric, we restrict ourselves to the setting of special relativity theory.
However, we expect that all the steps can be repeated for a general non-flat metric 
which corresponds to solving the electromagnetic field equations in a background gravitational field.
We approach the discretization of Maxwell's equations directly in space-time.
The choice of GA as a mathematical language comes from the fact that it enables concise and abstract notation,
which in turn lets us focus on introduced mathematical structures rather than their particular implementations.

\section{Space-Time Algebra (STA)}
\label{sec:STA}
The overview:
In this section we introduce the basic concepts from relativity theory and GA,
which are necessary for the paper.
We discuss the concept of physical and reference time, as well as what do we mean by an observer.
This concepts are crucial for the proper physical interpretation of the quantities involved.
We then introduce GA, STA and some concepts from geometric calculus,
which are used later.

To properly define the concept of the observer we need some basic notions, which are presented next.

The family of 1D curves filling densely, continuously and without intersections the space-time region
is called a congruence.
If the tangent vector $u$ to a curve (belonging to the congruence)
passing through any point is timelike, then the congruence is called timelike.
Time-like congruence may be interpreted as follows.
Each member of a family represents a world-line of a particle.
Particles can be imagined to form a body made of elastic rubber, which can deform freely but cannot tear apart.
As a consequence particles whose distance is small have similar velocity.

The congruence is a purely geometrical concept; there was no notion of time so far.
Since each particle of the congruence can move with different velocity, the time passes at different rates for each of them.
The time measured by a particle is called the proper time $\tau$ of that particle; we will also refer to it as the physical time.
In simple words, if a particle looks at its watch carried on its hand it sees $\tau$.

The proper time $\tau$ is a natural parametrisation of any member of the time-like congruence.
Of course, in the definition of the natural parameter there is a norm (associated with the metric) involved.
In our case, the metric is not positive definite (which is sometimes called pseudometric), thus
resulting in the pseudonorm (not always greater than zero).
However, since we have restricted ourselves to time-like curves,
the pseudonorm of the tangent vector $u$ is always positive, thus we can treat it as a usual norm,
and thus the proper time $\tau$ being a natural parameter is well defined (up to a constant as in the case of
the positive definite metric and the norm associated with it).
The reference time $t_{\text{ref}}$ is a parameter (not necessarily a natural one) running along
each member of the congruence.
It is required that if we join the points of different curves associated with the same value of $t_{\text{ref}}$
the obtained hypersurface, called hypersurface of simultaneity or constant time, is continuous.

With a time-like congruence there is associated the four-velocity field $u(x)$, with $x$ the position in space-time,
of tangent vectors. The congruence can be reconstructed from the four-velocity field by means of integration.
The integration constants are fixed by the particular choice of time synchronisation.

The whole structure: time synchronisation and the four-velocity field $u(x)$ (thus also the time-like congruence)
is referred to as the observer or the frame of reference.

The GA can be defined as a quotient algebra obtained by the factorisation of the full tensor algebra with
respect to a certain ideal, see \cite{fecko}.
However, we follow the approach presented in \cite[Chapter 4.1]{GAForPhys}.
Namely, the resulting properties of the product in quotient algebra are treated as axioms defining the geometric product.

The geometric product is
\begin{align}
AB &\neq BA \,, & \tag{noncommutative} \\
(AB)C&=A(BC)\,, & \tag{associative} \\
A(B+C)&=AB+AC \,, & \tag{left-distributive}\\
(A+B)C&=AC+BC\,, & \tag{right-distributive} 
\end{align}
where $A$, $B$ and $C$ are generic elements of GA called multivectors.
If $a$ is a vector, then its square is a real number
\begin{equation}
 a^2 = g(a,a) \in \mathbb{R} \,,
\end{equation}
with $g$ the metric tensor.

The geometric product of two vectors can be split into its symmetric and antisymmetric part
\begin{equation}
ab=\frac{1}{2}\left(ab+ba\right)+\frac{1}{2}\left(ab-ba\right):= a\cdot b+a\wedge b \,,
\end{equation}
where $\cdot$ and $\wedge$ denote the \textit{scalar} and \textit{exterior} product.

From the axioms follows that the scalar product is a real number
\begin{equation}
a\cdot b = \frac{1}{2}\left(ab+ba\right) = g(a,b)
\end{equation}
and coincides with the scalar product in linear algebra.

The exterior product $a\wedge b$ of two vectors is called bivector or $2$-vector.
It can be interpreted geometrically as an oriented surface of undefined shape and position, spanned by the vectors $a$ and $b$.
Thus a circle in one point and a rectangle at the other point, but with the same
area, given by $\left|(a\wedge b)^2\right|$, and parallel to each other are associated with the same bivector.

Any multivector that can be written as an exterior product of $n$ vectors
\begin{equation}
 V_n = v_1\wedge \dots \wedge v_n
\end{equation}
is called an $n$-blade.\\
Any sum of $n$-blades
\begin{equation}
 A_n = \sum\limits_{i} V_n^i
\end{equation}
is called an $n$-vector.

Let $d$ be the dimension of the space.
There are only two unit $d$-vectors, say, $I_1$ and $I_2$, related via $I_2 = -I_1$.
By unit we mean that $\left| I_i^2 \right| = 1$, with $i=1,2$.
$I_1$ and $I_2$ correspond to two possible orientations of the space.
The chosen unit $d$-vector is denoted by $I$ and called the pseudoscalar.
Once an ordered set of basis vectors $\gamma_k,\,k=1,\dots,d$ is given, the pseudoscalar is uniquely defined via
\begin{equation}
 I := \frac{\gamma_1\wedge\dots\wedge\gamma_d}{\sqrt{\left|\left(\gamma_1\wedge\dots\wedge\gamma_d\right)^2\right|}}\,.
\end{equation}

Since we are interested in $4$-dimensional space-time, we introduce more concepts particular for this setting.

Basis vectors are denoted by $\gamma_a$, with $a=0,1,2,3$.
The reciprocal basis vectors are denoted by $\gamma^a$ and defined by requiring that they satisfy 
$\gamma_a \cdot \gamma^b = \delta_a^b$, with $\delta_a^b$ the Kronecker delta.

The basis vectors of orthogonal Cartesian coordinate system $t,x,y,z$ are denoted by $\gamma_i$, with $i=t,x,y,z$.
The reciprocal basis is $\gamma^t = \gamma_t$ and $\gamma^\alpha = -\gamma_\alpha$, with $\alpha = x,y,z$.
This basis is orthonormal by definition, thus we will exploit basic properties such as
$\gamma_i \wedge \gamma_j = \gamma_i\gamma_j = - \gamma_j\gamma_i$ without explicitly commenting on that.

The pseudoscalar is given by
\begin{equation}
 I = 
 \frac{\gamma_0\wedge\gamma_1\wedge\gamma_2\wedge\gamma_3}
 {\sqrt{\left|\left(\gamma_0\wedge\gamma_1\wedge\gamma_2\wedge\gamma_3\right)^2\right|}}=
 \gamma_t\gamma_x\gamma_y\gamma_z \,.
\end{equation}

The 3D basis
\begin{equation}
 \sigma_\alpha := \hat{\gamma}_\alpha \wedge \hat{\gamma}_0
 \label{sigma}
\end{equation}
with $\hat{\gamma}_i := \gamma_i / \sqrt{|\gamma_i^2|}$,
is used to embed 3D vectors in the space-time setting.
This is also called space-time split 
(first introduced in \cite[Chapter 7]{HestenesPhD}, but the name appeared later in publications by the same author)
with respect to the observer with the four-velocity $\hat{\gamma}_0$.

By an arrow over a symbol we denote 3D vectors.
For example, the electric field
\begin{equation}
 \vec{E} := \sum\limits_{\alpha=1}^{3} E^\alpha \sigma_\alpha \,.
 \label{vectors}
\end{equation}
We would like to justify the name ``3D vectors''.
From the point of view of 4D algebra \eqref{sigma} are bivectors.
However, one can build a 3D (not necessarily geometric) algebra treating $\sigma_\alpha$ as basis vectors.
As a consequence, one can identify $\sigma_\alpha$ with 3D basis vectors from traditional vector analysis.
Therefore, traditional 3D vector calculus can be used in a relative space of any observer, provided
one uses \eqref{sigma} and \eqref{vectors} to embed 3D vectors in 4D algebra.

The electromagnetic field in space-time is described by the Faraday bivector
\begin{equation}
 F := \vec{E} + I \vec{B} \,.
\end{equation}
Knowing $F$ one can split it into electric and magnetic field observed in the reference frame with four-velocity $u$
\begin{align}
 \vec{E} &= \frac{1}{2} \left( F -u F u\right) \label{splitE} \\
 \vec{B} &= \frac{1}{2I} \left( F + u F u\right) \,. \label{splitB}
\end{align}
This split is determined solely by the four-velocity field of the observer.

The geometric derivative
\begin{equation}
 \nabla := \sum\limits_{i=0}^{3}\gamma^i \partial_i = 
 \gamma_t \partial_t - \gamma_x \partial_x - \gamma_y \partial_y - \gamma_z \partial_z
\end{equation}
is a vector from the algebraic point of view, and an operator acting on the object to the right of it from
the operational perspective.
In the same manner we define the 3D geometric derivative
\begin{equation}
 \vec{\nabla} := \sum\limits_{\alpha=1}^{3} \sigma_\alpha \partial_\alpha \,.
\end{equation}

For the comprehensive derivations and definitions of geometric calculus see 
\cite{Sobczyk}
or \cite{ShapeOfDiffGeo}, \cite[Chapter~6]{GAForPhys}, \cite{CAtoGC}.
As with the geometric product, we focus rather on a technical definition, i.e., how to evaluate integrals.

The geometric differential is defined as
\begin{equation}
 d^nx := e_1 \wedge \dots \wedge e_n dx^1 \dots dx^n \,,
\end{equation}
where $dx^i$ are scalar differentials and $e_i$ are unit vectors spanning the hypersurface,
thus $e_1 \wedge \dots \wedge e_n$ is a pseudoscalar of the hypersurface $K^n$, compare \cite[Equation~(3.4)]{Sobczyk}.

We define the integral of a multivector field A
\begin{equation}
 \int\limits_{K^n} d^nx A := \int\limits_{K^n} ( e_1 \wedge \dots \wedge e_n ) A dx^1\dots dx^n
\end{equation}
as the limit of a sum as is usually done in traditional setting
\footnote{
In general, the integrand $(e_1 \wedge \dots \wedge e_n) A$ is not a scalar, thus one needs to
define how to transport non-scalar quantities (in order to sum them up),
which in turn depends on the connection employed.
In this paper we work with flat Minkowski space-time, therefore this issue is irrelevant (the connection is trivial).
}.

If the integrand is a scalar then it can be written as
\begin{equation}
 (e_1 \wedge \dots \wedge e_n) \cdot A_n \,,
\end{equation}
and the integral
\begin{equation}
 \int\limits_{K^n} (d^nx) \cdot A_n
\end{equation}
can be related to the integral of a differential form.

\section{Maxwell's Equations in STA}
The overview:
Here we start from traditional 3D form of Maxwell's equations and constitutive laws and
derive their space-time equivalents expressed using GA.
This is our starting point for performing space-time discretization.

The traditional form of Maxwell's equations separating space and time reads
 \begin{align}
 \bigg\{ \begin{array}{r l}
  \vec{\nabla} \times \vec{E} &= -\partial_t \vec{B}\\
 \vec{\nabla} \cdotp \vec{B} &= 0  
 \end{array} 
 &&
 \bigg\{ \begin{array}{r l}
  \vec{\nabla} \times \vec{H} &= \partial_t \vec{D} + \vec{J} \\ 
 \vec{\nabla} \cdotp \vec{D} &= \varrho \,,
 \end{array}
 \label{ME3D}
 \end{align} 
where $\varrho$, $\vec{J}$, $\vec{E}$, $\vec{H}$, $\vec{D}$ and $\vec{B}$ 
are
the electric charge density, 
the electric current density,
the electric and magnetic field strengths, 
and the electric and magnetic flux densities, respectively.
It should be stressed that Maxwell's equations take the form \eqref{ME3D}
only in an inertial reference frame.

Maxwell's equations \eqref{ME3D} do not form a complete set of equations.
They need to be supplemented with constitutive material laws.
Since we focus on lossless wave propagation, 
the material laws in the reference frame, in which the materials are at rest,
read
\begin{align}
\vec{D} = \varepsilon \vec{E} && 
\mbox{and} && 
\vec{H} = \nu \vec{B} \,,
\label{material3D}
\end{align}
with $\varepsilon$ the electric permittivity and $\nu$ the magnetic reluctivity.
We assume that the medium is local and linear, possibly inhomogeneous and anisotropic.
From mathematical point of view this is equivalent to assuming that $\varepsilon$ and $\nu$
are positive-definite, second rank tensor fields.

Equations \eqref{ME3D} and \eqref{material3D} 
(with given boundary conditions and initial values) 
form a complete set of equations, which we want to solve.
Since we aim at performing space-time discretization, we must express \eqref{ME3D}--\eqref{material3D} in the space-time setting.

To this end,
we combine all 3D scalar and vector fields into space-time counterparts in the following way
\begin{align}
 F  := \vec{E} + c I \vec{B} && G  := \vec{D} + \frac{1}{c} I \vec{H} && J  :=  (c \varrho +\vec{J}) \gamma_0 \,,
 \label{FGJ}
\end{align}
where $\gamma_0$ is the four-velocity of the observer and $c$ is the speed of light in vacuum.

Thus, Maxwell's equations in the space-time form \cite{GAForPhys} ensue, viz.
\begin{align}
  \nabla \wedge F = 0 \,, &&
  \nabla \cdot  G = J  \,.
  \label{ME4D}
\end{align}

The space-time integral form of Maxwell's equations is obtained in a manner described in \cite{DFinGC}.
Namely, we integrate \eqref{ME4D} over an arbitrary bounded
3D star domain $K_3 \subset K_4$, with $K_4$ the 4D space-time,
and 
apply the fundamental theorem of geometric calculus 
\cite{DFinGC,Sobczyk}.
This results in (SI units in square brackets)
\begin{align}
 \oint\limits_{\partial K_3} (d^2x) \cdot F = 0 \quad \left[ V \cdot m \right] \,,
 &&
 \oint\limits_{\partial K_3} ( d^2x ) \wedge G = \int\limits_{K_3} (d^3x) \wedge J \quad \left[ C \cdot \frac{m}{s} \right] \,. 
 \label{ME4Dint}
\end{align}

The material mapping $\xi$ is implicitly defined through
\begin{equation}
G = \xi(F) \,.
\end{equation}
If the materials are isotropic, 
then $\varepsilon$ and $\nu$ are scalars, and we may easily obtain an explicit formula for $\xi$, viz.
(for a discussion on $\xi$ for anisotropic materials, the reader is referred to \cite{myHiroshima})
\begin{multline}
 G = \vec{D}+\frac{1}{c}I\vec{H} = \varepsilon \vec{E}+\frac{\nu}{c} I\vec{B} 
 = \varepsilon\frac{1}{2}\left( F-\gamma_0 F \gamma_0 \right) + \frac{\nu}{c^2} 
\frac{1}{2}\left( F+\gamma_0 F \gamma_0 \right) 
= \\=
\frac{1}{2}\left[ \left(\varepsilon +\frac{\nu}{c^2}\right) F - \left(\varepsilon - \frac{\nu}{c^2}\right) \gamma_0 F \gamma_0
\right] \,,
\end{multline}
i.e.,
\begin{equation}
\xi(X) = \frac{1}{2}\left[ \left(\varepsilon +\frac{\nu}{c^2}\right) X - \left(\varepsilon - \frac{\nu}{c^2}\right) \gamma_0 X \gamma_0 \right] \,.
\end{equation}
In vacuum, we have that $\varepsilon=\varepsilon_0$, $\nu=\nu_0$ and $c=\sqrt{\frac{\nu_0}{\varepsilon_0}}$,
from which the following simple mapping ensues
\begin{equation}
G = \xi(F) = \varepsilon_0 F \,.
\end{equation}

\section{Discretization}
The overview:
We first present how we construct the 4D mesh complex from 3D primal mesh.
It is convenient due to the fact that there is a wide variety of available 3D mesh generators
and the material distribution is constant in time with respect to the 3D mesh.
Then the integral form of Maxwell's equations is applied to this complex in order to discretize them.
We then introduce the interpolating functions; namely, Whitney ones.
They are later used in creation of material matrix, that is the discrete equivalent of the material mapping $\xi$.

In this section we use GA as well as linear algebra.
Therefore, we introduce the following notation.
The elements of linear algebra (vectors and matrices) are denoted with a bar under the symbol
\begin{equation}
 \underline{f} := \left[ f_1,\,f_2,\dots ,\, f_n \right] \,,
\end{equation}
where $n$ follows from the context (for example, in the case of $\underline{f}$ it is the number of 2D facets).
Analogously $\underline{M}$ is the matrix with entries $M_{ij}$.
If the quantities from linear algebra are multiplied, the product from that algebra is employed.

\subsection{4D Mesh Construction}
We introduce the placement map, which specifies the observer.
We derive the explicit expressions for the case of rotation with constant angular velocity.
This map is used to create the 4D primal mesh from a 3D one.
In the end we introduce two possible choices for the barycentric dual.
We want to stress that, one can use the barycentric dual mesh
(with broken edges)
in FIT as well.
We present, however, the FIT derivations for a dual with straight edges, as that is implemented in the code used to simulate
numerical examples.

\subsubsection{Placement Map}

We extend the Lagrangian description of motion to our space-time setting by defining the placement map
\begin{equation}
 p : T \times K_3 \to K_4 ,\, (t_\text{ref},\vec{r}_\text{ref} ) \mapsto r \,,
\end{equation}
where $t_{\text{ref}}$ is the reference time, $\vec{r}_{\text{ref}}$ is the reference position, and
$r$ is the physical position vector in space-time
\footnote{
The reference domain is meshed using any 3D mesh generator;
we obtain coordinates of the nodes in the form $\left[ r^x,r^y,r^z \right]$,
then $\vec{r}_\text{ref} := r^x\sigma_x + r^y\sigma_y + r^z\sigma_z $.
}.

Our description may be thought of as a map $ r_{\text{ref}} \mapsto r $ between reference and physical 4D positions.
Nonetheless, we explicitly separate the reference time $t_{\text{ref}}$ 
\begin{equation}
p_{t_\text{ref}} (\vec{r}_\text{ref}) := p(t_\text{ref}, \vec{r}_\text{ref} ) \,,
\end{equation}
because this enhances later explanations.

When comparing our placement map with the traditional Lagrangian description, we observe that
\begin{enumerate}
 \item We do not restrict the reference position $\vec{r}_{\text{ref}}$ to be
 the position at time zero. One reason for this is that there is no unique notion of time (as discussed in Section~\ref{sec:STA}).
 \item We do not restrict the time component $t$ of the position vector $r$ to be the same as the reference time $t_{\text{ref}}$.
 This allows to impose various synchronization of the clocks, i.e., change hypersurfaces of simultaneity.
\end{enumerate}

\subsubsection{Rotation with Constant Angular Velocity $\Omega$}

In order to deal with rotation we must define what we mean by constant angular velocity $\Omega$.
(This discussion can be performed analogously to the discussion of constant acceleration in the relativistic setting,
see \cite[Chapter 6.2]{MTW})
If one applies the constantly accelerated motion as $v = a t$, then after a long enough time, one gets $v>c$, 
which is obviously not consistent with relativity.
Therefore, one rather takes as definition that the velocity at a time instant $t + dt$ is $a dt$ greater,
with a relativistic formula for the addition of velocities, i.e., 
\begin{equation}
 v = \frac{v_1+v_2}{1+v_1 v_2 / c^2} 
\end{equation}
employed.
This corresponds to stating that the acceleration is constant as observed by the accelerated observer.
We do the same for constant rotation, that is,
we define that the velocity $v$ at the radius $r+dr$ is $\Omega dr$ higher than at $r$.
(The reader familiar with the derivation for accelerated observer will see that
from mathematical point of view we just replace $t$ with $r$
and $a$ with $\Omega$.)

Then at the radius $r+\Delta r$, we have that
\begin{align}
v(r+\Delta r) &= v(r) + \frac{dv}{dr} \Delta r + \mathcal{O}(\Delta r)^2 \\
v(r+\Delta r) &= \frac{v(r)+\Omega \Delta r}{1+v(r)\Omega \Delta r /c^2} = 
v(r) +\Omega \Delta r - \frac{v^2(r)}{c^2} \Omega\Delta r + \mathcal{O}(\Delta r)^2\,,
\end{align}
that is, in the limit $\Delta r \rightarrow 0$, the velocity $v$ satisfies the following ordinary differential equation
\begin{equation}
\frac{dv}{dr} = \Omega \left(1 - \frac{v^2}{c^2}\right) \,.
\end{equation}
By imposing the condition that the velocity at the center of rotation is zero $v(0)=0$, we arrive at the solution
\begin{equation}
v(r) = c \tanh\left(\frac{r\Omega}{c}\right) \,.
\label{constantRotation}
\end{equation}
For small radii, \eqref{constantRotation} gives $v(r) \approx r\Omega$, which corresponds to the Galilean setting.

The cylindrical coordinates of $\vec{r}_{\text{ref}}$ are denoted by $[r, \varphi, z]$.
For a system rotating with constant angular velocity $\Omega$ around $z$-axis we have
\begin{equation}
 p = p_t(\vec{r}_{\text{ref}}) =
 [t,r\cos(\theta),r\sin(\theta),z][\gamma_t, \gamma_x, \gamma_y, \gamma_z]^T
 \mbox{ , with } \theta = \varphi + \tanh\left(\frac{r\Omega}{c}\right) \frac{c t}{r} \,.
 \label{placement}
\end{equation}
Above, $p_t$ is such that it satisfies
$t_{\text{ref}}=t$ and $\vec{r}_{\text{ref}} = p_0(\vec{r}_{\text{ref}}) \wedge \gamma_t$
coinciding thus with 3D Lagrangian description of motion.
Moreover,
one can verify that the four velocity $u$ of a particle with reference coordinates $[r,\varphi,z]$ is consistent with
\eqref{constantRotation}, see Appendix~\ref{app:velocity}.

\subsubsection{Extrusion of the 3D Mesh}
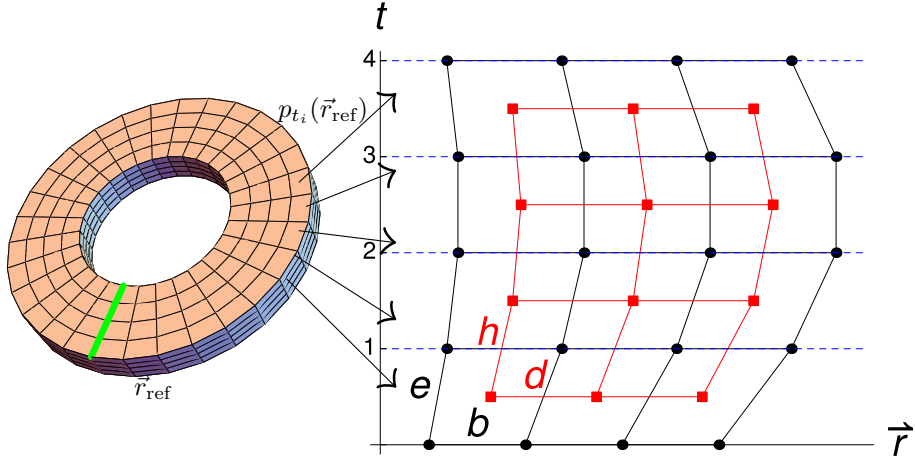
\begin{figure}
\begin{tikzpicture}[scale=.32]
\node[inner sep=0pt]  at (0,0)
    {\includegraphics[
    width=.43\linewidth]{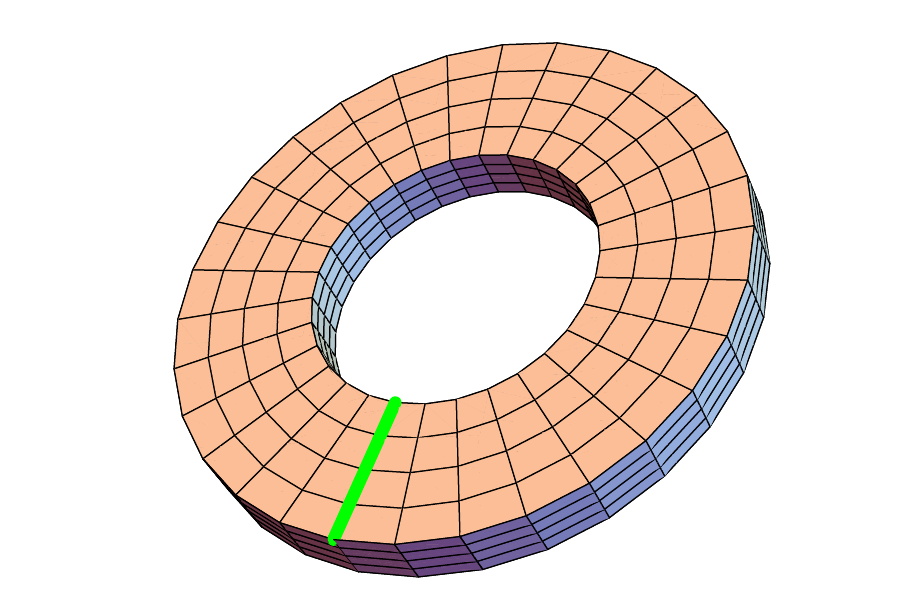}};
\node[inner sep=0pt]  at (20,0)
    {\includegraphics[width=.5\linewidth]{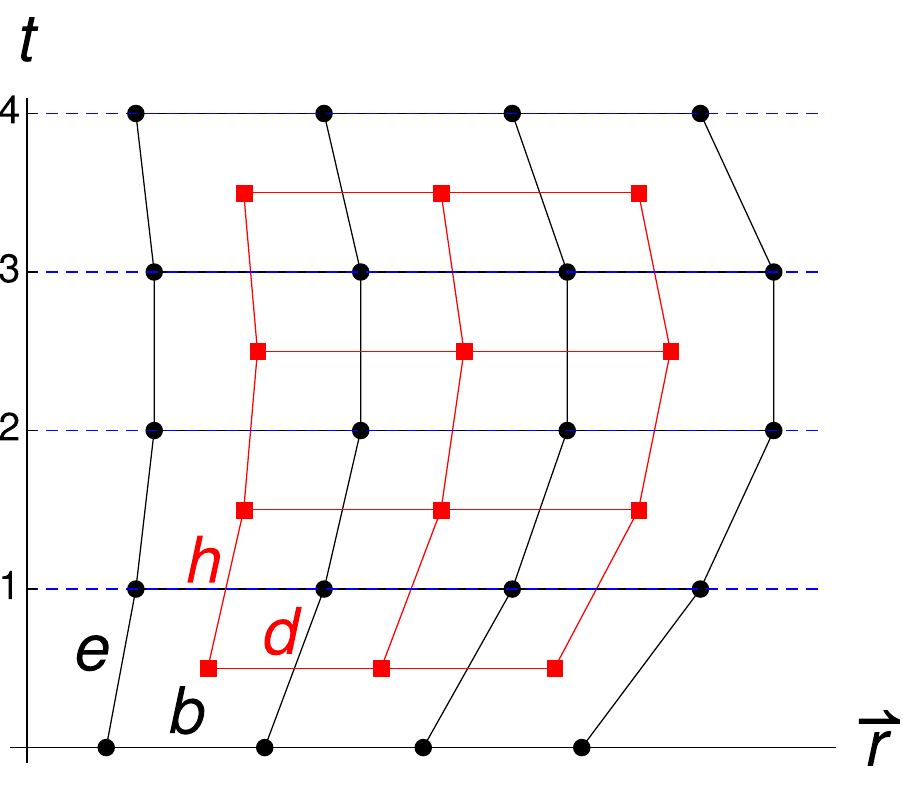}};
    \draw[
    decoration={markings,mark=at position 1 with {\arrow[scale=3]{>}}},
    postaction={decorate},
    shorten >=0.4pt
    ]
    (6.0,0.0) -- (10.0,-0.5);
    \draw[
    decoration={markings,mark=at position 1 with {\arrow[scale=3]{>}}},
    postaction={decorate},
    shorten >=0.4pt
    ]
    (6.3,1.0) -- (10.0,2.5);
        \draw[
    decoration={markings,mark=at position 1 with {\arrow[scale=3]{>}}},
    postaction={decorate},
    shorten >=0.4pt
    ]
    (6.0,2.0) -- (10.0,5.7);
    \draw[
    decoration={markings,mark=at position 1 with {\arrow[scale=3]{>}}},
    postaction={decorate},
    shorten >=0.4pt
    ]
    (5.8,-1) -- (10.0,-3.7);
    \draw[
    decoration={markings,mark=at position 1 with {\arrow[scale=3]{>}}},
    postaction={decorate},
    shorten >=0.4pt
    ]
    (5.5,-2) -- (10.0,-6.5);
\node[inner sep=0pt]  at (7,5.0) {$p_{t_i}(\vec{r}_{\text{ref}})$};
\node[inner sep=0pt]  at (0.0,-6.5) {$\vec{r}_{\text{ref}}$};
\end{tikzpicture}
    \caption{
    Sketch of a space-time mesh used in simulation.
    }
    \label{stmesh}
\end{figure}

We depict the mesh construction in Fig.~\ref{stmesh}.
Let $M^3$ be a 3D reference mesh depicted as a toroid in the leftmost of Fig.~\ref{stmesh}.

We proceed to create images of $M^3$ under the placement map $p_t(M^3)$ at times $t = t_1,t_2,...,t_i$.
This is readily carried out by applying $p_t$ to the position vector $\vec{r}_{\text{ref}}$ of the nodes in $M^3$ while
preserving the connectivity, i.e., the incidence relations between elements.
The resulting space-time nodes are depicted as black points in Fig.~\ref{stmesh}
and the elements inherited from $M^3$ as black horizontal lines joining these points.
We then connect, along the ``vertical'' time line (see Fig.~\ref{stmesh}), the nodes corresponding to
the same reference point in $M^3$.
The resulting 4D mesh is referred to as $M^4$, and we refer to any $n$-dimensional element of $M^4$ as $K_n^i$.

\subsubsection{Two Choices for Barycentric Dual}
We define two kinds of dual meshes;
namely,
one is $\bar{M}^4$ that will be used in a 4D FIT context, and the other is $\widetilde{M}^4$ in a 4D FEM context.
In both cases the connectivity and dual nodes are the same.
As depicted in Fig.~\ref{stmesh} with red squares, the dual nodes are the barycenters of the 4D cells of $M^4$.
Therefore, it appears natural to label a dual node within a 4D cell $K_4^i$ with the same index $i$. 
Nodes are used to form higher dimensional elements in the following way.
If $K_4^i \cap K_4^j = K_3^k$ then
the dual nodes $i$ and $j$ are connected, and the resulting connecting edge is labelled with index $k$.
In the same frame of ideas,
the dual edge $k$ belongs to the dual facet $l$ if $K_2^l \subset K_3^k$,
and so on for dual 3D and 4D elements.

Although we have defined the connectivity of all elements and positions of the dual nodes,
the exact geometry of the dual elements is not settled by these definitions.
In fact,
there are two cases in which we are interested and describe them next.

\begin{figure}
\centering
\begin{tikzpicture}[scale=1.5]
  \coordinate [label={below left:$\widetilde{K}_0^{i_1}$}] (n1) at (-1, -1);
  \coordinate [label={below right:$\widetilde{K}_0^{i_2}$}] (n2) at (1, -1);
  \coordinate [label={above right:$\widetilde{K}_0^{i_3}$}] (n3) at (1, 1);
  \coordinate [label={above left:$\widetilde{K}_0^{i_4}$}] (n4) at (-1, 1);
  \coordinate [label={below:$b$}] (b) at (0,0);
  
  \draw (n1) -- (n2) -- (n3) -- (n4) -- (n1);
  \draw (n1) -- (b);
  \draw (n2) -- (b);
  \draw (n3) -- (b);
  \draw (n4) -- (b);
  
  \node[inner sep=0pt]  at (0,-.7) {$T_1$};
  \node[inner sep=0pt]  at (0,.7) {$T_3$};
  \node[inner sep=0pt]  at (.7,0) {$T_2$};
  \node[inner sep=0pt]  at (-.7,0) {$T_4$};
\end{tikzpicture}
\caption{
A dual facet of $\bar{M}^4$ with $n=4$ edges. The barycenter of the dual nodes $\widetilde{K}_0^{i}$ is denoted by $b$.
The facet consists of $n=4$ flat triangles $T_i$.
}
\label{dualFacet}
\end{figure}
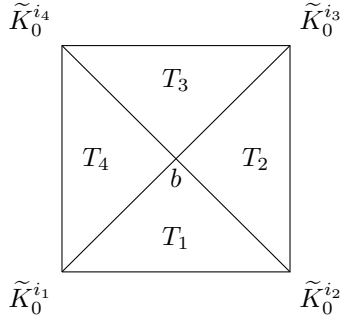

The barycentric dual with straight edges $\bar{M}^4$ is obtained by joining the dual nodes
with a straight line segment.
The dual facet with $n$ edges (see Fig.~\ref{dualFacet}) is composed of $n$ flat triangles with the nodes at dual edge's nodes plus the barycenter 
of all dual nodes belonging to the dual facet.
If the dual nodes are coplanar our definition reduces to the flat dual facet.
The dual 3D element is defined as bounded by the dual facets.

\begin{figure}
 \centering
 \begin{tikzpicture}[place/.style={circle,fill=black,inner sep= 0pt,minimum size = .3em}]
  \draw (-2,-1) -- (2,-1) -- (2,0) -- (0,1) -- (-2,0) -- (-2,-1);
  \draw [line width = .2em, red ] (0,-1) -- (0,1);
  
  \node at (-1,-0.25) [place] {};
  \node at (+1,-0.25) [place] {};
  \node at (0,0) [place] {};
  
  \coordinate [label={above right:$x_j^\text{b}$}] (b) at (0,0);
  \coordinate [label={below right:$x_j^\text{i}$}] (b) at (0,-.25);
  
  \coordinate [label={above left: \textcolor{red}{$K^j_2$}}] (b) at (0,0);
  \coordinate [label={below right: \textcolor{blue}{$\widetilde{K}^j_2$}}] (b) at (-1,-0.25);
  
  \draw [dashed, blue] (-1,-0.25) -- (+1,-0.25);
  \draw [dotted, blue] (-1,-0.25) -- (0,0) -- (+1,-0.25);
 \end{tikzpicture}
 \caption{
 The difference between dual edges of $\bar{M}^4$ (dashed line) and $\widetilde{M}^4$ (dotted line).
 The intersection point $x_j^\text{i}$ of $K_2^j$ and $\widetilde{K}_2^j$ in $\bar{M}^4$ is
 not the barycenter $x_j^\text{b}$ of $K_2^j$ as in $\widetilde{M}^4$.
 Nevertheless, we later use the interpolated fields at $x_j^\text{b}$ in our extension of FIT material matrices.
 }
 \label{fig:dualEdges}
\end{figure}
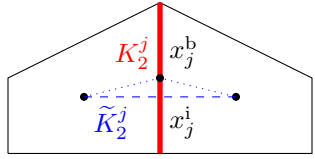

The barycentric dual $\widetilde{M}^4$ is defined as the 4D extension of the 3D construction
described in \cite[page 99]{BossavitBook}.
Briefly speaking, the dual edge is defined as a broken line segment consisting of
the dual nodes and the barycenter of a corresponding 3D element (see Fig.~\ref{fig:dualEdges}).
This construction is extended to higher dimensional dual elements in such a way
that the intersection of the corresponding dual and primal objects is always the barycenter of the primal one.
Moreover, each dual object is a sum of flat simplices.
This construction in 2D and 3D case is depicted in \cite[Fig. 4.3 and Fig. 4.4]{BossavitBook}, respectively.

Now, let us assume that the reference mesh $M^3$ bears some orientation of facets and edges.
To uniquely define the orientation of the 4D mesh $M^4$ we adopt the convention ``first time then space''.
Therefore, if the edge is represented by a vector $p$, then the timelike facet associated with that edge has the orientation
given by the product $u\wedge p$.

\textbf{Remark:}
In our space-time extension of FIT material matrices we apply several heuristics
such as the one explained in the caption of Fig.~\ref{fig:dualEdges}.
In brief, one may perceive 4D FIT as potentially more efficient, but there is no underlying theory, e.g.,
error analysis, i.e., convergence proof.
On the other hand 4D FEM is built on the sound mathematical framework being, however, computationally more expensive.
Therefore, we find it interesting to present both approaches to the discretization of material mapping.

\subsection{Maxwell's Grid Equations}
\label{sec:gridEquations}
The overview:
Here we discretize Maxwell's equations.
After general discretization in space-time,
we introduce some further simplifications coming from our particular construction of 4D mesh.
The physical interpretation of DoFs coming from these simplifications is given,
and relation to the discrete boundary conditions is briefly discussed.
Our goal was to stick to the 3D FIT with leapfrog time integration, as close as possible.

We discretize Maxwell's equations by applying their integral form \eqref{ME4Dint} to the primal-dual mesh pair.
The \eqref{ME4Dint}-left is applied to the primal grid and \eqref{ME4Dint}-right to the dual one.
This yields
\begin{align}
  \oint\limits_{\partial K_3^i} (d^2x) \cdot F = 0 \,,
 &&
 \oint\limits_{\partial \widetilde{K}_3^i} ( d^2x ) \wedge G = \int\limits_{\widetilde{K}_3^i} (d^3x) \wedge J \,. 
\end{align}
Please note that integrals over the boundary of primal and dual 3-cells $\partial K_3^i$ and $\partial \widetilde{K}_3^i$
are sums of integrals over 2D facets, i.e.,
\begin{align}
 \oint\limits_{\partial K_3^i} (d^2x) \cdot F = \sum\limits_j \int\limits_{K_2^j} (d^2x) \cdot F \,,
 &&
 \oint\limits_{\partial \widetilde{K}_3^i} ( d^2x ) \wedge G = \sum\limits_j \int\limits_{\widetilde{K}_2^j} ( d^2x ) \wedge G \,.
\end{align}
Therefore,
in the case of a general mesh, like a simplicial 4D one as used in \cite{simplices4D},
we would introduce scalar DoFs
\begin{align}
 f_j := \int\limits_{K_2^j} (d^2x) \cdot F \,,
 &&
 g_j := I^{-1} \int\limits_{\widetilde{K}_2^j} ( d^2x ) \wedge G \,,
 &&
 j_i := I^{-1}\int\limits_{\widetilde{K}_3^i} (d^3x) \wedge J \,,
 \label{DoFs}
\end{align}
obtaining thus the discrete Maxwell's equations
\begin{align}
 \underline{D}^2 \underline{f} = 0 \,,
 &&
 \underline{\widetilde{D}}^2 \underline{g} = \underline{j} \,,
 \label{ME4Dgrid}
\end{align}
where $\underline{D}^2$ and $\underline{\widetilde{D}}^2$ are 
incidence matrices describing adjacency of 2D to 3D elements of the primal and dual mesh,
respectively.
Matrices $\underline{D}^n$ have entries
\begin{equation}
 D^n_{ij} :=
 \begin{cases}
  +1 \text{ if } K_n^j \subset K_{n+1}^i \text{ with matching orientation } \\
  -1 \text{ if } K_n^j \subset K_{n+1}^i \text{ with opposite orientation } \\
  0 \text{ if } K_n^j \not\subset K_{n+1}^i
 \end{cases}
 \,.
\end{equation}

Nevertheless,
we have chosen a structured 4D mesh, and we shall exploit this fact in order to simplify the discrete form of Maxwell's equations.
Since in this paper we are interested in wave propagation phenomena, we assume $J=0$ in \eqref{FGJ} to simplify our derivations.
This assumption does not come from the limitations of the method.

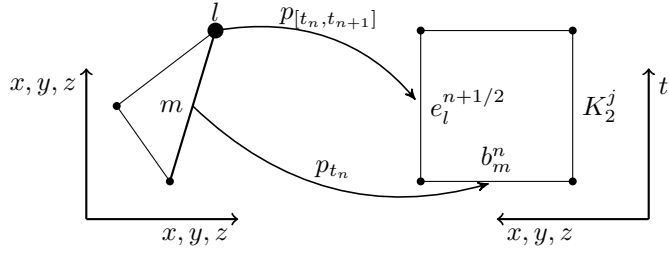
\begin{figure}
\centering
\begin{tikzpicture}[place/.style={circle,fill=black,inner sep= 0pt,minimum size = .3em}, scale=2]
  \coordinate [label={left:$m$}] (facet) at (0, 0);
  \coordinate [label={above:$l$}] (edge1) at (.15, .5);
  \coordinate (edge2) at (-.15, -.5);
  \coordinate (edge3) at (-.5, 0);
  \coordinate [label={right:$e^{n+1/2}_l$}] (e) at (1.5, 0);
  \coordinate [label={above:$b^n_m$}] (b) at (2, -.5);
  \coordinate [label={right:$K_2^j$}] (K) at (2.5,0);
  
  \coordinate (c1) at (1.5,-.5);
  \coordinate (c2) at (2.5,-.5);
  \coordinate (c3) at (2.5,+.5);
  \coordinate (c4) at (1.5,+.5);
  
  \node at (edge1) [style={circle,fill=black,inner sep= 0pt,minimum size = .6em}] {};
  \node at (edge2) [place] {};
  \node at (edge3) [place] {};
  \draw [thick] (edge1) -- (edge2);
  \draw (edge1) -- (edge3) -- (edge2);
  
  \draw (c1) -- (c2) -- (c3) -- (c4) -- (c1);
  \node at (c1) [place] {};
  \node at (c2) [place] {};
  \node at (c3) [place] {};
  \node at (c4) [place] {};
  
  \path (edge1) edge [bend left,->,>=stealth',shorten >=3pt,semithick] node [above] {$p_{\left[ t_n,t_{n+1} \right]}$} (e);
  \path (facet) edge [bend right,->,>=stealth',shorten >=3pt,semithick] node [above] {$p_{t_n}$} (b);
  
  \coordinate (f0) at (-.7 , -.75);
  \coordinate (fx) at (.3 , -.75);
  \coordinate (fy) at (-.7, .25);
  
  \draw[thick,->] (f0) -- (fx) node[anchor=north east] {$x,y,z$};
  \draw[thick,->] (f0) -- (fy) node[anchor=north east] {$x,y,z$};
  
  \coordinate (f20) at (3 , -.75);
  \coordinate (f2x) at (2 , -.75);
  \coordinate (f2y) at (3, .25);
  
  \draw[thick,->] (f20) -- (f2x) node[anchor=north west] {$x,y,z$};
  \draw[thick,->] (f20) -- (f2y) node[anchor=north west] {$t$};
  
\end{tikzpicture}
\caption{
Illustration of the relation between indices in \eqref{DoFsSplit}.
The leftmost is the reference mesh. The $l$-th edge is depicted as a fat dot, and $m$-th 2D facet as a thick line.
The rightmost is the space-time mesh.
2D facets are depicted as lines.
}
\label{fig:indices}
\end{figure}

Next, we rename and renumber the DoFs on the primal mesh as follows
\begin{equation}
  f_j = \int\limits_{K_2^j} (d^2x) \cdot F =: 
 \begin{cases}
  e^{n+1/2}_l & \mbox{if $K_2^j$ is timelike} \,,\\
  b^{n}_m & \mbox{if $K_2^j$ is spacelike} \,,
 \end{cases}
 \label{DoFsSplit}
\end{equation} 
where the relation between the indices $j$, $l$, $m$ and $n$ is as follows (see also Fig.~\ref{fig:indices}).
If the facet $K_2^j$ is timelike, 
it is the edge $l$ in the reference mesh $M^3$
extruded in time via the placement map $p_t$ such that $t \in \left[t_n,t_{n+1} \right]$.
Similarly, if the facet $K_2^j$ is spacelike,
then it is the image of a facet with index $m$ in $M^3$
under the map $p_t$ with $t=t_n$.

We perform a similar decomposition of the dual DoFs
\begin{equation}
 g_j = I^{-1} \int\limits_{\widetilde{K}_2^j} ( d^2x ) \wedge G =:
 \begin{cases}
  d^{n+1/2}_l & \mbox{if $\widetilde{K}_2^j$ is spacelike} \\
  h^{n}_m & \mbox{if $\widetilde{K}_2^j$ is timelike}
 \end{cases}
 \,.
\end{equation}

In simple words, instead of an arbitrary enumeration of the 2D facets in $M^4$,
we rather label them by using the facet/edge numbering in $M^3$ and the time step to which they correspond.
This numbering scheme is always possible and unique due to our particular construction of the 4D mesh.
Such a decomposition aids the physical interpretation of the DoFs,
as well as
enables a simple time marching procedure,
which is in general not possible on unstructured 4D meshes as considered in \cite{simplices4D}.

We now move towards the physical interpretation of the previously introduced DoFs.
By $\vec{E}_\text{s}$ and $\vec{B}_\text{s}$ we denote the 3D fields obtained by performing
the space-time split by a stationary observer, i.e., 
the one with the four-velocity $u = \gamma_t$ in \eqref{splitE} and \eqref{splitB}.
Analogously, the fields $\vec{E}_\text{r}$ and $\vec{B}_\text{r}$ are associated with (the 
four-velocity $u$ given by \eqref{urot} of) the rotating observer.

\paragraph{Spacelike case:}
Let us consider a spacelike facet $K_2^j$,
and let us suppose that it is spanned by orthonormal vectors $\gamma_1$ and $\gamma_2$
and
parametrized with local coordinates $x^1$ and $x^2$.
The ``normal'' vector $\gamma_3$ is uniquely defined by requiring that ${\gamma_i}$, with $i=t,1,2,3$
form an orthonormal set with the orientation $\gamma_t \gamma_1\gamma_2\gamma_3 = +I$.
Then
\begin{multline}
 b^{n}_m =
 \int\limits_{K_2^j} (d^2x) \cdot F = 
 \int\limits_{K_2^j} \left((\gamma_1dx^1)\wedge(\gamma_2dx^2) \right) \cdot (\vec{E}_\text{s} + I \vec{B}_\text{s}) = \\ =
 \int\limits_{K_2^j} dx^1dx^2 (\gamma_1\gamma_2) \cdot (E_\text{s}^i \gamma_i\gamma_t + I B_\text{s}^i\gamma_i\gamma_t) =
 \int\limits_{K_2^j} dx^1dx^2 B_\text{s}^3 \,.
\end{multline}
Now we bring the expression above to the form from traditional 3D vector analysis.
First, we note that vector surface element $d\vec{S} :=\vec{n} dS$, where the normal vector $\vec{n}$ is
defined as a cross product of vectors spanning the surface; that is
\begin{equation}
 \vec{n} := \sigma_1 \times \sigma_2 = -I \sigma_1 \wedge \sigma_2 = \sigma_3 \,.
\end{equation}
Next, the magnetic field component may be written as
\begin{equation}
 B_\text{s}^3 = \sigma_3 \cdot \vec{B}_\text{s} = \vec{n} \cdot \vec{B}_\text{s} \,.
\end{equation}
Therefore, in general it holds
\begin{equation}
 b^{n}_m = 
 \int\limits_{K_2^j} dx^1dx^2 B_\text{s}^3 =
 \int\limits_{K_2^j} dS (\vec{n} \cdot \vec{B}_\text{s}) =
 \int\limits_{K_2^j} d\vec{S} \cdot \vec{B}_\text{s} \,.
\end{equation}

\paragraph{Timelike case:}
Now, let us consider a timelike facet $K_2^j$, and let us
suppose that it is spanned by orthonormal vectors $u$ and $p$.
The local coordinates of the facet are: the proper time $\tau$ and the distance along the edge $l$.
Then we get analogously 
\begin{equation}
 e^{n+1/2}_l =
 \int\limits_{K_2^j} (d^2x) \cdot F =
 \int\limits_{K_2^j} \left((u \, d\tau)\wedge(p \, dl) \right) \cdot (\vec{E}_\text{r} + I \vec{B}_\text{r}) =
 - \int\limits_{K_2^j} d\tau \, dl \, \vec{p} \cdot \vec{E}_\text{r} \,,
\end{equation}
where $\vec{p} := p\wedge u$ is a 3D vector representing an edge.

The same reasoning leads to the physical interpretation of the dual DoFs $d$ and $h$.
The DoFs denoted by $e$ and $h$ are related to the electric and magnetic fields $\vec{E}_\text{r}$ and $\vec{H}_\text{r}$,
respectively, in the moving reference frame,
While
the DoFs denoted by $d$ and $b$ are related to the electric and magnetic flux densities $\vec{D}_\text{s}$ and $\vec{B}_\text{s}$,
respectively, in the stationary reference frame.

The physical interpretation of the DoFs enhances the imposition of boundary conditions.
For example, if we consider stationary Perfect Electric Conductor (PEC) boundary conditions,
then in 3D formulation one sets $e=0$ on the boundary, which corresponds to setting the tangential component
of $\vec{E}_\text{s}$ to zero, i.e.,
\begin{equation}
 \vec{n}_\text{s} \times \vec{E}_\text{s} = 0 \,,
\end{equation}
where $\vec{n}_\text{s}$ is the vector normal to PEC in a stationary reference frame.

If the PEC is rotating, then one needs to set the tangential component of $\vec{E}_\text{r}$
to zero at the boundary, i.e.,
\begin{equation}
 \vec{n}_\text{r} \times \vec{E}_\text{r} = 0 \,,
 \label{PECr}
\end{equation}
where $\vec{n}_\text{r}$ is the vector normal to PEC in a rotating reference frame.
Nevertheless, due to the particular construction of the mesh and the resulting physical interpretation of
the DoFs, it is clear that \eqref{PECr} is equivalent to setting $e=0$ on the boundary.
Therefore, from the implementation point of view nothing has changed due to movement.

Using \eqref{DoFsSplit} gives another advantage, namely, allows for a straightforward time marching procedure.
The Maxwell's grid equations \eqref{ME4Dgrid}
(with omitted discrete equivalents of $\mathrm{div} \vec{B}_\text{s} = 0$ and $\mathrm{div} \vec{D}_\text{s} = \varrho_\text{s}$)
now read
\begin{align}
 \underline{b}^{n+1} = \underline{b}^{n} + \underline{C} \, \underline{e}^{n+1/2} \,,
 &&
 \underline{d}^{n+1/2} = \underline{d}^{n-1/2} +\widetilde{\underline{C}} \, \underline{h}^{n} \,,
 \label{ME4DgridEBDH}
\end{align}
which are first solved for $n=1$, then for $n=2$, and so on.
$\underline{C}$ and $\widetilde{\underline{C}}$ \cite{Weiland1996} are, respectively, primal and dual facets to edges incidence matrices of the reference 3D mesh,
that is discrete curl operators.

The signs in \eqref{ME4DgridEBDH} depend on our convention ``first time then space'' for orientations of elements.
The convention affects the decomposition of, e.g., $\underline{D}^2$, in \eqref{ME4Dgrid} to the temporal
difference and 3D curl-matrix $\underline{C}$.
For relation of our DoFs to the FIT ones, see \eqref{FITDoFs}.

We want to point out the similarity with 3D FIT + leapfrog \cite{Weiland1996}.
It was shown, e.g., in \cite{myHiroshima,Mattiussi}, that leapfrog
comes from the topology of the primal/dual mesh pair.
However, this is only true if the material matrices are local in time, that is
they relate only the fields at the same time step, i.e.,
\begin{align}
 \underline{d}^{n+1/2} = \underline{M_{\varepsilon e}} \underline{e}^{n+1/2} \,, 
 && 
 \underline{h}^{n} = \underline{M_{\nu b}} \underline{b}^{n}
 \,.
\end{align}
Nevertheless, Maxwell's grid equations, i.e., topological laws 
\cite[Equation~(4)]{Hehl},
have not changed, because the movement is encoded only in the geometry of the mesh, not in the topology.
However,
our scheme does not always reduce to leapfrog as we consider space-time setting.
Material matrices may relate DoFs at different time steps,
which in turn results in changing the resulting time integration scheme.

\subsection{Whitney Interpolating Functions}
The overview:
The concept of Whitney elements is explained in GA notation.
Then we present them for a particular reference element.
The transformation to the physical domain is performed later.
The special properties of the lowest-order basis functions
are discussed in the reference element as well as in the physical ones,
that is the images of the reference element with respect to the particular diffeomorphism we use.

\subsubsection{General Introduction}
With every $n$-vector field $A_n$ there is associated its discrete equivalent $n$-cochain $\underline{A}_n$,
which is a vector in linear algebra sense, whose components are given by
\begin{equation}
A_n^i := \int\limits_{K_n^i} (d^nx) \cdot A_n \,,
\end{equation}
where $K_n^i$ is the $i$-th $n$-dimensional cell of the mesh.

The deRham complex
\begin{equation}
\begin{CD}
A_0  @> \nabla\wedge >>
A_1  @> \nabla\wedge >>
A_2  @> \nabla\wedge >>
A_3  @> \nabla\wedge >>
A_4   
\\
@A W^0 AA 
@A W^1 AA
@A W^2 AA
@A W^3 AA
@A W^4 AA
\\
\underline{A}_0  @> D^0 >>
\underline{A}_1  @> D^1 >>
\underline{A}_2  @> D^2 >>
\underline{A}_3  @> D^3 >>
\underline{A}_4 
\end{CD}
\label{deRhamComplex}
\end{equation}
will only commute for a proper choice of basis functions, namely,
those belonging to the class of Whitney elements.
In general there are infinitely many members of this class.
However, we restrict our attention to the lowest-order elements,
and thus they are uniquely defined up to a sign coming from the orientation of the facets in $\Xi$.
Therefore, in the sequel by Whitney elements, we mean lowest-order Whitney elements.
For lowest-order elements one can define Whitney reconstruction operator (Whitney map) as
\begin{equation}
W^n \underline{A}_n = \sum\limits_i A_n^i N^n_i(x) \,,
\label{WhitneyReconstruction}
\end{equation}
with $N^n_i(x)$ the $n$-vector valued basis function associated with $n$-cell $K_n^i$.

\subsubsection{Our Choice of Whitney Forms on the Reference Element}
So far, we have not specified any particular form of basis functions $N^n_{i}(x)$.
Due to the fact that we intend to obtain a computational scheme, the explicit formulas for Whitney functions $N^n_{i}(x)$ are necessary.
For the class of meshes discussed in this paper, 4D elements are tensor products of 3D elements and 1D temporal intervals.
Therefore, the extension of any particular 3D Whitney elements to 4D ones is straightforward.
Since in our implementation we work with hexahedral 3D mesh, our 4D elements are ``4D hexahedrons''.
Thus we shall extend here the 3D reasoning described in \cite[Chapter 8.3.]{FEM} to 4D.

We start by constructing the interpolating functions, that is Whitney elements, 
for a single tesseract (``4D reference cube'') $\Xi = [-1,+1] \otimes[-1,+1] \otimes[-1,+1] \otimes[-1,+1]$.
The vertices of this tesseract are at the points $(t,x,y,z)=(\pm1,\pm1,\pm1,\pm1)$.
Let us suppose we are interested in the basis function associated with the facet $[-1,+1]\otimes [-1,+1]\otimes -1 \otimes -1$.
Since this is a $tx$-facet, 
we know that the DoF living on such a facet is related to the $(\gamma_t\gamma_x) \cdot F = E^x$ 
component of the Faraday field $F = \vec{E}+I\vec{B}$.
Clearly the function
\begin{equation}
 N^f_1=\frac{1}{16} \gamma ^x \gamma ^t (1-y) (1-z) 
\end{equation}
vanishes at the other $tx$-facets and is orthogonal to the remaining facets, thus
\begin{equation}
 \int\limits_{\Delta_i} (d^2x)\cdot N^f_1 = \delta^i_1 \,,
\end{equation}
where $\Delta_i$, with $i=1,\dots,24$ are 2D facets of the reference tesseract.

One may construct a basis functions for all 2D facets, such that they fulfil the so-called interpolation property
\begin{equation}
 \int\limits_{\Delta_i} (d^2x)\cdot N^f_j = \delta^i_j \,,\quad i,j =1,\dots, 24,
 \label{normalization}
\end{equation}
and give an exact interpolation for constant fields.
Moreover, these basis functions guarantee that the tangential component of the interpolated field $F$ is continuous
across a 2D facet, viz.
\begin{equation}
 \widetilde{W} \cdot F_1 = \widetilde{W} \cdot F_2 \,,
 \label{continuity}
\end{equation}
where $F_1$ and $F_2$ are the limits of the field $F$ as approached from any two directions.
Mathematically speaking
\begin{equation}
 F_i = \lim_{\delta \to 0} F(x + \delta v_i) \,,\, i=1,2,
\end{equation}
where $v_i$ is a unit vector describing the direction.

The space-time split of \eqref{continuity} leads to the more familiar 3D continuity relations
\begin{align}
 \vec{B}_1 \cdot \vec{n} = \vec{B}_2 \cdot \vec{n}  && \vec{E}_1 \cdot \vec{l} = \vec{E}_2 \cdot \vec{l} \,,
\end{align}
with $\vec{n}$ and $\vec{l}$, respectively, a normal and a tangent vector to an arbitrary surface.

Regarding the physical interpretation
of $N^2_{i\pm,\,j\pm}$ (defined in \eqref{Whitney2D}) in the case of $\Phi_1$ being an identity map, 
Table~\ref{tab:interpretation} provides a complete description.
\begin{table}
\centering
\begin{tabular}{c|cccccc}
 $(i,j)$ & $(x,t)$ & $(y,t)$ & $(z,t)$ & $(x,y)$ & $(z,x)$ & $(y,z)$ \\ \hline
 $\text{Field component}$ & $B^x$ & $B^y$ & $B^z$ & $E^z$ & $E^y$ & $E^x$
\end{tabular}
\caption{The physical interpretation of DoFs associated with the basis functions $N^2_{i\pm,\,j\pm}$.}
\label{tab:interpretation}
\end{table}

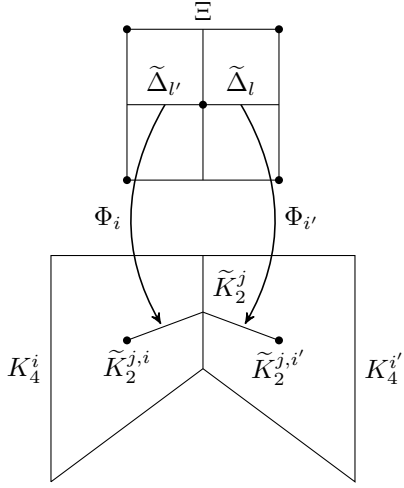
\begin{figure}
\centering
\begin{tikzpicture}[place/.style={circle,fill=black,inner sep= 0pt,minimum size = .3em}, scale=2]
  \coordinate (n1) at (-1, -2.5);
  \coordinate (n2) at (0,-1.75);
  \coordinate (n3) at (1,-2.5);
  \coordinate (n4) at (1,-1);
  \coordinate (n5) at (0,-1);
  \coordinate (n6) at (-1,-1);
  
  \coordinate [label={above right:$\widetilde{K}_2^j$}] (w0) at (0,-1.375);
  \coordinate [label={below :$\widetilde{K}_2^{j,i}$}] (w1) at (-.5, -1.5625);
  \coordinate [label={below :$\widetilde{K}_2^{j,i'}$}] (w2) at (+.5, -1.5625);

  \coordinate (c1) at (-.5,-.5);
  \coordinate (c2) at (+.5,-.5);
  \coordinate (c3) at (+.5,+.5);
  \coordinate (c4) at (-.5,+.5);
  
  \coordinate (b0) at (0,0);
  \coordinate (b1) at (0,-.5);
  \coordinate (b2) at (.5,0);
  \coordinate (b3) at (0,.5);
  \coordinate (b4) at (-.5,0);

  \draw (c1) -- (c2) -- (c3) -- node [above] {$\Xi$} (c4) -- (c1);
  \node at (c1) [place] {};
  \node at (c2) [place] {};
  \node at (c3) [place] {};
  \node at (c4) [place] {};

  \node at (b0) [place] {};
  \draw (b0) -- (b1);
  \draw (b0) -- node [above] {$\widetilde{\Delta}_l$} (b2);
  \draw (b0) -- (b3);
  \draw (b0) -- node [above] {$\widetilde{\Delta}_{l'}$} (b4);

  \draw (n1) -- (n2) -- (n3) -- node [right] {$K_4^{i'}$} (n4) -- (n5) -- (n6) -- node [left] {$K_4^{i}$} (n1);
  \draw (n2) -- (n5);
  
  \node at (w1) [place] {};
  \node at (w2) [place] {};
  \draw (w1) -- (w0) -- (w2);
  
  \path (-.25,0) edge [bend right,->,>=stealth',shorten >=3pt,semithick] node [left] {$\Phi_{i}$} (-.25,-1.5);
  \path (+.25,0) edge [bend left,->,>=stealth',shorten >=3pt,semithick] node [right] {$\Phi_{i'}$} (+.25,-1.5);
  
\end{tikzpicture}
\caption{
The facets $\widetilde{\Delta}_{l'}$ and $\widetilde{\Delta}_l$ of the reference tesseract $\Xi$ are mapped
via $\Phi_{i}$ and $\Phi_{i'}$to the parts 
$\widetilde{K}_2^{j,i}$ and $\widetilde{K}_2^{j,i'}$ of dual facets $\widetilde{K}_2^j$
contained in $K_4^i$ and $K_4^{i'}$, respectively.
Namely 
$\Phi_{i}\left( \widetilde{\Delta}_{l'} \right) = \widetilde{K}_2^{j,i}$
and
$\Phi_{i'}\left( \widetilde{\Delta}_{l} \right) = \widetilde{K}_2^{j,i'}$.
The bivector associated with 
$\widetilde{K}_2^j = \widetilde{K}_2^{j,i} \cup \widetilde{K}_2^{j,i'}$ 
is thus
$\widetilde{W}_j = \widetilde{W}_{j,i} + \widetilde{W}_{j,i'}$.
}
\label{fig:consistencyCondition}
\end{figure}

We observe that the chosen basis functions $N^2_i$, besides being Whitney forms (see Appendix~\ref{sec:ExplicitWhitney}),
do also comply with the requirements of an energetic approach 
(as introduced in \cite{EnergeticApproach}) in our space-time extension
\cite[equation iii)]{Codecasa},
that is
\begin{equation}
 \int\limits_{\Xi} N^2_j |d^4x| = I^{-1} \widetilde{W}_j \,,
 \label{energeticCondition}
\end{equation}
where $\Xi$ is the reference tesseract and $\widetilde{W}_j$ is the bivector associated with the facet $\widetilde{\Delta}_j$
dual to $\Delta_j$,
which is only the part contained in the reference tesseract $\Xi$. 
Please note that the dual facets $\widetilde{K}^j_2$ in the whole mesh complex are sums of the facets
$\Phi_i(\widetilde{\Delta}_j)$
in neighbouring 4D cells $K_4^i$.
Therefore, the bivectors associated with $\widetilde{K}_2^{j,i}$, being the part of the dual facet $\widetilde{K}^j_2$
contained in the primal cell $K_4^i$,
are later referred as $\widetilde{W}_{j,i}$.
This explanation is sketched schematically in Fig.~\ref{fig:consistencyCondition}.

The property \eqref{energeticCondition} transfers to the physical domain as follows.
Suppose it holds in the reference domain.
In this reasoning the quantities in the reference domain are denoted by a bar.

First, we note that
\begin{align}
 \int\limits_{K^i_4} N_j |d^4x| &= \lim_{\Delta^4 x \to 0} \sum\limits_{a, \, x_a \in K^i_4} N_j(x_a) |\Delta^4 x_a| = \\
 &= \lim_{\Delta^4 \bar{x} \to 0} \sum\limits_{a, \, \bar{x}_a \in \Xi}
 \Phi_i\left( \bar{N}_j (\bar{x}_a) \right) \Phi_i\left(|\Delta^4 \bar{x}_a|\right) = \\
 &= \lim_{\Delta^4 \bar{x} \to 0} \sum\limits_{a, \, \bar{x}_a \in \Xi}
 \Phi_i\left( \bar{N}_j (\bar{x}_a) |\Delta^4 \bar{x}_a|\right) = \\
 &= \Phi_i\left( \lim_{\Delta^4 \bar{x} \to 0} \sum\limits_{a, \, \bar{x}_a \in \Xi}
  \bar{N}_j (\bar{x}_a) |\Delta^4 \bar{x}_a|\right) 
 = \Phi_i\left( \int\limits_{\Xi} \bar{N}_j |d^4\bar{x}| \right) \,,
\end{align}
where $|\Delta^4x|$ is a finite 4D volume element and we assumed that interchanging
the order of diffeomorphism and the limit of the sum is justified.
Moreover, we have assumed that both the reference and the physical domain are equipped with the flat Minkowski metric,
thus the connection is trivial and is not necessary in the sum.

So we calculate in physical domain
\begin{multline}
 \int\limits_{K^i_4} N_j |d^4x| = 
 \Phi_i\left( \int\limits_{\Xi} \bar{N}_j |{d^4\bar{x}}| \right) =
 \text{ use \eqref{energeticCondition} } =
 \Phi_i\left(\bar{I}^{-1} \bar{\widetilde{W}} \right) = \\ =
 \Phi_i\left(\bar{I}^{-1}\right) \Phi_i\left( \bar{\widetilde{W}} \right) =
 I^{-1} \Phi_i\left( \bar{\widetilde{W}} \right) \,,
\end{multline}
where $\Phi$ is the mapping to the physical domain,
and we used the fact that the unit pseudoscalar is invariant of the mapping ($I$ was defined so).
Therefore \eqref{energeticCondition} will hold in physical domain if
\begin{equation}
 \Phi\left( \bar{\widetilde{W}} \right) = \widetilde{W} \,,
\end{equation}
which will hold if $\Phi$ maps the reference dual facet to the physical dual facet.
Therefore,
$\Phi \in \text{affine mapping} \Rightarrow
\Phi\left( \bar{\widetilde{W}} \right) = \widetilde{W} \Rightarrow
\text{ \eqref{energeticCondition} }$.
So if the mapping preserves (weighted) barycenters then \eqref{energeticCondition} is implied, i.e.,
one cannot came up with a map that preserves barycenters and violates \eqref{energeticCondition}.

\subsubsection{Basis Functions in the Physical Domain}

The basis functions $N^n_i$ satisfy the following set of properties:
\begin{enumerate}
 \item they satisfy the interpolation property \eqref{normalization},
 \item they fulfil \eqref{WhitneyCondition}, i.e.,
 guarantee that \eqref{deRhamComplex} commutes (see Appendix~\ref{sec:ExplicitWhitney}),
 \item they are eligible for 4D energetic approach, i.e., meet \eqref{energeticCondition}.
\end{enumerate}
All these properties hold for the reference tesseract;
however, in many applications one is interested in using meshes with non-cubic elements
in order to approximate geometries more accurately.
Moreover, if one even uses a 3D Cartesian grid, the 4D elements are certainly deformed by the movement, 
which is encoded in the geometry of the 4D mesh.
Therefore, one needs to construct analogous functions for a deformed tesseract.
This is conveniently done by first defining a mapping (diffeomorphism) from the reference to the physical tesseract
and then by transforming the basis functions accordingly.

We define a set of mappings
\begin{equation}
 \Phi_p : \Xi \rightarrow K_4^p 
\end{equation}
from the reference domain to the physical 4D cells.
One of important features of these mappings is that they are affine maps, consequence of which is
that the barycenters of any object in the reference domain are mapped to 
the barycenters of a corresponding object in the physical domain.
For details see Appendix~\ref{sec:Phi}.

\begin{figure}
\centering
\begin{tikzpicture}[place/.style={circle,fill=black,inner sep= 0pt,minimum size = .3em}, scale=2]
  \coordinate (n1) at (1.3,-.45);
  \coordinate (n2) at (+2.7,-.65);
  \coordinate (n3) at (+2.3,+.7);
  \coordinate (n4) at (1.75,+.4);
  
  \coordinate (s0) at (2.0125,0);
  \coordinate (s1) at (2., -0.55);
  \coordinate (s2) at (2.5, 0.025);
  \coordinate (s3) at (2.025, 0.55);
  \coordinate (s4) at (1.525, -0.025);
  
  \coordinate (biased) at (2.4, 0.25);
  
  \coordinate (c1) at (-.5,-.5);
  \coordinate (c2) at (+.5,-.5);
  \coordinate (c3) at (+.5,+.5);
  \coordinate (c4) at (-.5,+.5);
  
  \coordinate (b0) at (0,0);
  \coordinate (b1) at (0,-.5);
  \coordinate (b2) at (.5,0);
  \coordinate (b3) at (0,.5);
  \coordinate (b4) at (-.5,0);

  \draw (c1) -- (c2) -- (c3) -- node [above] {$\Xi$} (c4) -- (c1);
  \node at (c1) [place] {};
  \node at (c2) [place] {};
  \node at (c3) [place] {};
  \node at (c4) [place] {};

  \node at (b0) [place] {};
  \draw (b0) -- (b1);
  \draw (b0) -- node [below] {$\widetilde{\Delta}_l$} (b2);
  \draw (b0) -- (b3);
  \draw (b0) -- (b4);
  
  \draw (n1) -- (n2) -- (n3) -- node [above] {$K_4^i$} (n4) -- (n1);
  \node at (n1) [place] {};
  \node at (n2) [place] {};
  \node at (n3) [place] {};
  \node at (n4) [place] {};
  
  \node at (s0) [place] {};
  \draw (s0) -- (s1);
  \draw (s0) -- node [below] {$\widetilde{K}_2^{j,i}$} (s2);
  \draw (s0) -- (s3);
  \draw (s0) -- (s4);
  
  \draw [dashed] (s0) -- (biased);
  
  \path (+.25,0) edge [bend left,->,>=stealth',shorten >=3pt,semithick] node [above] {$\Phi_i$} (+2.25,.15);
  
\end{tikzpicture}
\caption{
The facets $\widetilde{\Delta}_l$ are mapped in general to $\Phi_p(\widetilde{\Delta}_l)$,
which does not necessarily coincide with the part of the dual facet $\widetilde{K}_2^j$.
However, for the mapping discussed in this paper $\Phi_p(\widetilde{\Delta}_l) = \widetilde{K}_2^j$.
}
\label{fig:barycenters}
\end{figure}
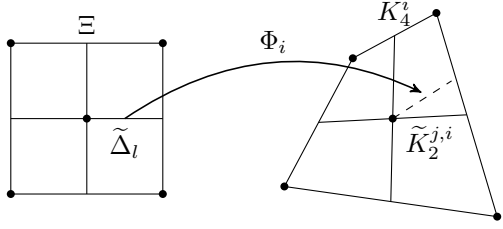

Property~1. involves the integral of a differential form, which is invariant with respect to the change of the metric properties;
therefore, this property will hold for the transformed functions.
As explained in \cite[text below Equation (2.13)]{Hiptmair},
the transformed (pulled-back) functions will satisfy the Property~2. on a deformed element.
Property~3. is not guaranteed in general as the dual elements of $\Xi$ are not mapped
to the barycentric dual elements of the deformed tesseract.
However, for the mapping we use, i.e., $\Phi_p$, this is the case, that is barycenters are mapped to barycenters, thus
the reference barycentric dual is mapped to the physical barycentric dual; see Fig.~\ref{fig:barycenters}.
Hence, all
required properties are inherited by the deformed tesseract.
Even if that was not the case, one can still use the energetic approach,
by simply stating that a different dual (not the barycentric one) is used.
Since in general, one does not need to calculate the dual mesh in FEM,
this bears no additional computational cost.
The price to pay is that one cannot easily relate $d$ and $h$ to $\vec{D}$ and $\vec{H}$
without calculating the dual.

In other words, properties 1. and 2. come from differential topology, thus change in the metric, i.e., shape of an element,
does not influence them.
In contrast, the property~3. depends on metric in a twofold way.
First, the definition of the barycenter depends on the notion of distance, thus the barycentric dual depends on metric.
Second, the integrand in \eqref{energeticCondition} is not a scalar; therefore, the Levi-Civita connection is used, which
depends on the metric.

There is no unique approach in the literature how to define the basis vectors in the physical domain. 
We opt for the following definitions
\begin{align}
 \bar{\gamma}_i := \frac{\partial x}{ \partial \bar{x}^i } \,, && \bar{\gamma}^i := \nabla \bar{x}^i(x) \,,
\end{align}
where $x$ is a position vector in physical space and $\bar{x}^i(x)$ are reference coordinates
(in $\bar{\gamma}_i$ basis)
of $x$, i.e.,
\begin{equation}
 \bar{x}^i = \bar{\gamma}^i \cdot \Phi^{-1}_p (x) \,.
\end{equation}

Then we easily verify that they ``transform'' according to
\begin{align}
 \bar{\gamma}_i &= \frac{\partial}{\partial \bar{x}^i} x =
 \frac{\partial x^j}{\partial \bar{x}^i} \frac{\partial}{\partial x^j} x =
 \frac{\partial x^j}{\partial \bar{x}^i} \gamma_j \,, \\
 \bar{\gamma}^i &= \nabla \bar{x}^i = 
 \left( \gamma^j \frac{\partial}{\partial x^j} \right) \bar{x}^i =
 \frac{\partial \bar{x}^i}{\partial x^j} \gamma^j \,.
\end{align}
It means that basis vectors transform like vectors, and the reciprocal basis vectors as one-forms.
It is less surprising if one realises that $\gamma_i \triangleq \partial_i$ and $\gamma^i \triangleq (dx^i)^\#$,
where the quantities on the right hand side are defined in exterior algebra of differential forms
(the canonical musical isomorphism is used, as we assume the existence of some metric).
The technical details of transforming $N^2_i$ are in the Appendix~\ref{sec:technicalTransformation}.

\subsection{Material Matrices}
The overview:
As we discussed before, the discretized Maxwell's equations are the same in FIT and Whitney FEM on the primal/dual cell complex.
From computational point of view, the only difference is in the discretization of material laws.
In this section we present our extension of FIT material matrices
and the energetic approach generalised to the space-time setting.
In both approaches the error is introduced in this step.
In the following we assume that, the material is allocated in primal grid,
i.e., in each 4D cell $K^i_4$ there is a constant material mapping $\xi_i$.

\subsubsection{4D FIT with Whitney Interpolation}
The overview:
Since we work with a non-orthogonal mesh pair, the traditional FIT approach for creating diagonal material matrices
is not possible.
The construction of FIT material matrices on non-orthogonal grids has been approached by several authors.
The closest approach to ours is described in \cite{Schuhmann},
where the construction of material matrices for (non-orthogonal) 2D quadrilateral meshes in FIT framework was proposed.
However, it is not clear how this work can be extended to space-time.
Thus, we rather use our general STA discretization and Whitney interpolation, which are ready to be applied in space-time.
We would like to mention that the sparsity pattern of the material matrices are exactly the same as in \cite{Schuhmann}
\footnote{
Of course, to compare the 3D method \cite{Schuhmann} and our 4D one, one needs to either
extend the first one to 4D or more straightforwardly reduce our scheme to 3D in the case of no rotation $\Omega=0$.}
and are sparser than in 4D energetic approach.
The error is introduced in \eqref{FITapprox} by assuming that the fields are constant on every dual facet
$\widetilde{K}_2^j$. 
This assumption is needed,
to recover the fields by knowing only the DoFs, i.e.,
integrals over $\widetilde{K}_2^j$.
Then $\xi$ can be used to relate the fields, resulting in the relation between DoFs.
This is done with an approximation error.

We recall cf.~\eqref{DoFs} that the dual DoFs are defined as
\begin{equation}
 I g_j = \int\limits_{\widetilde{K}_2^j} (d^2x)\wedge G = \int\limits_{\widetilde{K}_2^j} (d^2x)\wedge \xi(F) \,.
 \label{dualDef}
\end{equation}
In order to construct the material matrices, we express
$F$ in terms of the primal DoFs $f_j$ and use the formula above.

We follow the general FIT approach for approximating the integral in \eqref{dualDef};
namely, approximate it by taking the value of $F$ at only one point $x_j$.
It may be thus perceived as an extension of the one-point (midpoint?) quadrature rule to the geometric calculus.
Thus, we approximate \eqref{dualDef} by
\begin{equation}
 \int\limits_{\widetilde{K}_2^j} (d^2x)\wedge \xi(F) \approx
 \left[ \int\limits_{\widetilde{K}_2^j} (d^2x)\right] \wedge \overline{\xi}(F_j) = \widetilde{W}_j \wedge \overline{\xi}(F_j) \,,
 \label{FITapprox}
\end{equation}
where $F_j := F(x_j)$ should be the field at the intersection of $\widetilde{K}_2^j$ with $K_2^j$
(or the barycenter of $\widetilde{K}_2^j$ according to midpoint rule).
$\overline{\xi}$ is $\xi$ averaged over the dual facet $\widetilde{K}_2^j$.
However, due to the fact that it is easy to interpolate the field at the barycenter of the primal facet $K_2^j$
we have used that point, see Fig.~\ref{fig:dualEdges}.
Therefore, we are approximating the integral in \eqref{dualDef} by the value of $F$ at the point which is not necessary lying
on $\widetilde{K}_2^j$.

The mapping \eqref{diffeoStart}--\eqref{diffeoEnd} is easy to calculate, i.e.,
having given $(\bar{t},\bar{x},\bar{y},\bar{z})$ it is easy to calculate $(t,x,y,z)$.
However, it is difficult to invert, i.e.,
having given $(\bar{t},\bar{x},\bar{y},\bar{z})$ it is not possible to obtain an analytic expression for
$(t,x,y,z)$;
one needs to rather go for numerical techniques to do that.
Although it is difficult to invert it is easy to ``guess'' $(\bar{t},\bar{x},\bar{y},\bar{z})$ of some interesting points, e.g.,
barycenters.
Being more precise: barycenters are mapped to barycenters.
Therefore, if one is interested in the interpolated fields at the barycenter of $K_2^j$ which corresponds
to, e.g., 2D facet $\Delta_1$ in the reference tesseract
with vertices $(\pm1,\pm1,-1,-1)$,
then
its barycenter is given by $(\bar{t},\bar{x},\bar{y},\bar{z})=(0,0,-1,-1)$.
The same holds true for barycenters of edges, 2D facets, 3D volumes and the whole tesseract.

Using Whitney interpolation  
\begin{equation}
 F(x) = \sum\limits_{k} f_k N^2_k(x) \,,
\end{equation}
we calculate
\begin{multline}
 I g_j = 
 \int\limits_{\widetilde{K}_2^j} (d^2x)\wedge \xi(F) =
 \sum\limits_{i} \int\limits_{\widetilde{K}_2^{j,i}} (d^2x)\wedge \xi_i(F) \approx
 \sum\limits_{i} \widetilde{W}_{j,i} \wedge \xi_i\left(F\left(x_j\right)\right) = \\=
 \sum\limits_{i} \widetilde{W}_{j,i} \wedge \xi_i\left( \sum\limits_{k} f_k N^2_k(x_j)\right) =
 \sum\limits_{k} \left[ \sum\limits_{i} \widetilde{W}_{j,i} \wedge \xi_i\left( N^2_k(x_j)\right) \right] f_k
\end{multline}
and thus, we are led to the material matrix $M_\xi$
\begin{equation}
 \left[M_\xi\right]_{jk} := I^{-1} \sum\limits_{i} \widetilde{W}_{j,i} \wedge \xi_i\left( N^2_k(x_j)\right) \,.
\end{equation}
Above $\widetilde{K}_2^{j,i}$ is the part of $\widetilde{K}_2^j$ contained in $K_4^i$.

In the formula above the averaging is done explicitly.
Therefore, it might be used to define the averaged $\overline{\xi}_j$ in \eqref{FITapprox}
by requiring that it satisfies
\begin{equation}
 \widetilde{W}_j\wedge \overline{\xi}_j (F_j) = \sum\limits_{i} \widetilde{W}_{j,i} \wedge \xi_i(F_j) \,.
\end{equation}

Due to the simplicity of predicting reference position of barycenters, we have used $x_j := \text{barycenter of }\Delta_j$.
The choice of $x_j \in K_2^j$ does not influence the fact that
the obtained material matrix $M_\xi$ is non-symmetric and thus endangers stability.
However, this issue can be heuristically solved by taking only the symmetric part of $M_\xi$ as
the material matrix. 
The numerical experiments (without rotation) show that such operation does not influence the convergence of the scheme,
while making it stable.

\subsubsection{4D FEM}
The overview:
We extend here to space-time the approach in \cite{Codecasa}.
The error is introduced in \eqref{discreteAction1} by assuming that the field $G$ is constant in each 4D element $K_4^i$.

Maxwell's equations can be obtained by applying variational principle $\delta S =0$ to the action integral,
e.g., \cite[Sections 2.4 and 2.5]{Stern}
\begin{equation}
 S = \frac{1}{2}\int\limits_{K_4} F \cdot G |d^4x| \,.
 \label{action}
\end{equation}
Since we have already discretized the operators, we will use $S$ only to derive
discrete material equations.

We define the action integral with explicitly applied material laws
\begin{equation}
 S_F = \frac{1}{2}\int\limits_{K_4} F \cdot \xi(F) |d^4x| \,.
 \label{action_F}
\end{equation}

The core idea is to spot that
\begin{equation}
 G = 2 \frac{\delta S}{\delta F} = \frac{\delta S_F}{\delta F} = \xi(F) \,,
 \label{functionalDerivatives}
\end{equation}
where $\delta/\delta F$ is partial functional derivative with respect to $F$.
Indeed, the variation of $S$ with respect to $F$ is
\begin{equation}
 \delta S = \int\limits_{K_4} \frac{G}{2} \cdot \delta F |d^4x| \,,
\end{equation}
and the variation of $S_F$
\begin{equation}
 \delta S_F = 
 \frac{1}{2}\int\limits_{K_4} \delta F \cdot \xi(F) |d^4x| + \frac{1}{2}\int\limits_{K_4} F \cdot \xi(\delta F) |d^4x| =
 \int\limits_{K_4} \xi(F) \cdot \delta F |d^4x| \,,
\end{equation}
where we used the fact that the material mapping is self-conjugate, i.e., $ B_1 \cdot \xi(B_2) = \xi(B_1) \cdot B_2 $ for all
bivectors $B_1$ and $B_2$.

Here we extend the 3D reasoning found in \cite[Section III]{Codecasa} to our space-time setting.
Approximating $G$ by a piecewise constant field, i.e., in each element $K_4^i$ it has the value $G_i$
and
assuming the basis functions satisfy
\begin{equation}
 \int\limits_{K_4^i} N^2_j |d^4x| = I^{-1}\widetilde{W}_{j,i} 
 \quad \Rightarrow \quad
 \sum\limits_{i} G_i \cdot \int\limits_{K_4^i} N^2_j |d^4x| = g_j \,,
\end{equation}
we obtain (the approximation sign is due to using $G(x) \approx G_i$ if $x \in K_4^i$)
\begin{equation}
 S = \frac{1}{2} \sum\limits_{i,\,j} \int\limits_{K_4^i} f_j N^2_j \cdot G |d^4x| \approx
 \frac{1}{2} \sum\limits_{i,\,j} f_j  G_i \cdot \int\limits_{K_4^i} N^2_j |d^4x| =
 \frac{1}{2} \sum\limits_{j} f_j g_j \,,
 \label{discreteAction1}
\end{equation}
from which follows
\begin{equation}
 g_i \approx 2 \frac{\partial S}{\partial f_i} \,.
\end{equation}

The action integral reads
\begin{equation}
 S_F = \frac{1}{2} \int\limits_{K_4} F \cdot \xi(F) |d^4x| =
 \frac{1}{2} \sum\limits_{i,\,j} f_i f_j \int\limits_{K_4} N^2_i \cdot \xi(N^2_j) |d^4x| \,.
 \label{discreteAction2}
\end{equation}

Using $S_F = S$ and comparing action integrals in \eqref{discreteAction1} and \eqref{discreteAction2} one gets
\begin{equation}
 g_i \approx \sum\limits_{j} \left[ \int\limits_{K_4} N^2_i \cdot \xi(N^2_j) |d^4x| \right] f_j =: 
 \left[\underline{M}_\xi \underline{f}\right]_i \,.
\end{equation}

By this construction, we have translated \eqref{functionalDerivatives} to the discrete setting in an approximate sense
\begin{equation}
 g_i \approx 2 \frac{\partial S}{\partial f_i} = \frac{\partial S_F}{\partial f_i} = 
 \left[\underline{M}_\xi \underline{f}\right]_i \,.
\end{equation}

\section{Resulting Scheme}
The overview:
In this section, we perform linear algebraic manipulations to obtain the numerical scheme.
In FIT case, \eqref{updateScheme} may be seen as an extension of leapfrog to the rotating case or
material equations, for which $d$ and $h$ depend on $e$ and $b$ as well.
We have avoided any extrapolation w.r.t. time as it may be the cause of instabilities
as mentioned in \cite[Appendix]{Novitski} and confirmed by our numerical examples.

In general the decomposition of $ \underline{g} = \underline{M}_\xi \underline{f}$ according to $e,b,d,h$ would read
\begin{align}
 h^n &= M_{\nu b}^-(n)b^{n-1} + M_{\nu e}^-(n)e^{n-1/2} + M_{\nu b}(n) b^n +
 \nonumber \\ + &
 M_{\nu e}^+(n)e^{n+1/2} + M_{\nu b}^+(n)b^{n+1}
 \label{materialH} \\
 d^{n+1/2} &= M_{\varepsilon b}^-(n+1/2)b^{n} + M_{\varepsilon e}(n+1/2)e^{n+1/2} + M_{\varepsilon b}^+(n+1/2)b^{n+1} \,,
 \label{materialD}
\end{align}
where the material matrices are defined according to 
\begin{equation}
\hspace{-5em}
 \begin{bmatrix}
  h^0 \\ d^{1/2} \\ h^1 \\ d^{3/2} \\ h^2  \\ \vdots \\ h^{n-1} \\ d^{n-1/2} \\ h^{n}
 \end{bmatrix}
 =
 \begin{bmatrix}
  M_{\nu b} & M_{\nu e}^+ & M_{\nu b}^+ & 0 & 0 & \dots & 0 & 0 & 0  \\
  M_{\varepsilon b}^- & M_{\varepsilon e} & M_{\varepsilon b}^+ & 0 & 0 & \dots & 0 & 0 & 0  \\
  M_{\nu b}^- & M_{\nu e}^- & M_{\nu b} & M_{\nu e}^+ & M_{\nu b}^+ & \dots & 0 & 0 & 0  \\
  0 & 0 & M_{\varepsilon b}^- & M_{\varepsilon e} & M_{\varepsilon b}^+ &  \dots & 0 & 0 & 0  \\
  0 & 0 & M_{\nu b}^- & M_{\nu e}^- & M_{\nu b} &  \dots & 0 & 0 & 0  \\
  \vdots & \vdots & \vdots & \vdots & \vdots & \ddots & \vdots & \vdots & \vdots  \\
  0 & 0 & 0 & 0 & 0 &   \dots & M_{\nu b} & M_{\nu e}^+ & M_{\nu b}^+  \\
  0 & 0 & 0 & 0 & 0 &   \dots & M_{\varepsilon b}^- & M_{\varepsilon e} & M_{\varepsilon b}^+  \\
  0 & 0 & 0 & 0 & 0 &  \dots & M_{\nu b}^- & M_{\nu e}^- & M_{\nu b} 
 \end{bmatrix}
 \begin{bmatrix}
  b^0 \\ e^{1/2} \\ b^1 \\ e^{3/2} \\ b^2 \\ \vdots \\ b^{n-1} \\ e^{n-1/2} \\ b^{n}
 \end{bmatrix} \,.
\end{equation}
Being more explicit, the block $M_{\nu e}^+(n)$ is given by
\begin{equation}
 M_{\nu e}^+(n) := M_\xi( 
 n(n_f+n_e)+ n_f + 1 : (n+1)(n_f+n_e) ,
 n(n_f+n_e)+1 : n(n_f+n_e) + n_f 
 )
\end{equation}
where $n_f$ and $n_e$ are, respectively, the number of facets and edges in the reference mesh.
The other blocks are defined analogously.

In general all material matrices depend on the time step $n$.
However, if the angular velocity $\Omega$ is constant and we use a constant time step $\Delta t$,
then all the matrices are constant in time.
This comes from the fact that the 4D mesh has certain symmetry.
Namely, the layer of cells below (=before) and above (=after) time $n$ has the same geometry as
the layer corresponding to $n+1$.
Thus in what follows we skip the dependency on $n$ to simplify the notation.

We can recover 3D FIT with leapfrog time integration in our framework when the angular velocity $\Omega = 0 $.
The relation between our DoFs and the ones used in \cite{Weiland1996} is
(the signs come from the conventions employed in this paper)
\begin{align}
 d = -\bbow{d} &&
 b = \bbow{b}  &&
 e = - \Delta t \bow{e} &&
 h = - \Delta t \bow{h} \,.
 \label{FITDoFs}
\end{align}
Then the time step looks as follows
\begin{align}
h^n &= M_{\nu b} b^n \nonumber \\
d^{n+1/2} &= d^{n-1/2} + \widetilde{C} h^n \nonumber \\
e^{n+1/2} &= M_{\varepsilon e}^{-1} d^{n+1/2} \nonumber \\
b^{n+1} &= b^n + C e^{n+1/2} \,,
\end{align}
which is usually written in more compact form as
\begin{align}
 e^{n+1/2} &= e^{n-1/2} + M_{\varepsilon e}^{-1} \widetilde{C} M_{\nu b} b^n \\
 b^{n+1} &= b^n + C e^{n+1/2} \,.
\end{align}

The Maxwell's grid equations \eqref{ME4DgridEBDH} do not change with the geometry of the mesh.
Only material matrices are affected by this change.
Therefore, when $\Omega \neq 0$ then we obtain 
(the first and the third line are \eqref{materialH} and \eqref{materialD}, respectively)
\begin{align}
h^n &= M_{\nu b}^-b^{n-1} + M_{\nu e}^-e^{n-1/2} + M_{\nu b} b^n + M_{\nu e}^+e^{n+1/2} + M_{\nu b}^+b^{n+1} \label{magnetic} \\
d^{n+1/2} &= d^{n-1/2} + \widetilde{C} h^n \label{dual} \\
e^{n+1/2} &= M_{\varepsilon e}^{-1} \left[ d^{n+1/2} - M_{\varepsilon b}^-b^{n} - M_{\varepsilon b}^+b^{n+1}  \right] \label{electric} \\
b^{n+1} &= b^n + C e^{n+1/2} \,.
\label{primal}
\end{align}

For 4D FIT $M_{\nu b}^- = M_{\nu b}^+ = 0$.
This comes from the fact that the basis functions associated with $b^{n+1}$ and $b^{n-1}$ vanish at
the time step $n$.
Therefore, there is no contribution of $b^{n+1}$ and $b^{n-1}$ to $h^n$.
This is the manifestation of the fact that our extension of FIT has narrower stencil.

For 4D FEM symmetry of $M_\xi$ translates to the following symmetries of $M$'s
\begin{align}
 M_{\nu b} = M_{\nu b}^T &&
 M_{\varepsilon e} = M_{\varepsilon e}^T &&
 M_{\varepsilon b}^- = \left( M_{\nu e}^+ \right)^T &&
 M_{\varepsilon b}^+ = \left( M_{\nu e}^- \right)^T &&
 M_{\nu b}^- = \left( M_{\nu b}^+ \right)^T \,.
\end{align}

To keep the workflow, i.e., keep the time integrator scheme explicit,
as in stationary case, 
one would need to, e.g., extrapolate 
$e^{n+1/2}$ and $b^{n+1}$ in \eqref{magnetic} and \eqref{electric}.
This however, leads to instabilities as we will demonstrate later.

Therefore, we derive an implicit scheme.
This can be done by, e.g., by plugging \eqref{primal}, \eqref{dual} and \eqref{magnetic} into \eqref{electric}.
After some algebra one arrives at
\begin{align}
 e^{n+1/2} &= M^{-1} \left[ \gamma b^{n} + \beta e^{n-1/2} + \alpha b^{n-1} \right] \\
 b^{n+1} &= b^n + C e^{n+1/2} \,,
 \label{updateScheme}
\end{align}
where
\begin{align}
 M :=& M_{\varepsilon e} - \widetilde{C} M_{\nu e}^+ - \widetilde{C} M_{\nu b}^+ C + M_{\varepsilon b}^+ C \\
 \alpha :=& M_{\varepsilon b}^- + \widetilde{C} M_{\nu b}^- \\
 \beta :=& M_{\varepsilon e} + \widetilde{C} M_{\nu e}^- \\
 \gamma :=& \widetilde{C} M_{\nu b} + \widetilde{C} M_{\nu b}^+ - M_{\varepsilon b}^-
\end{align}
The good news are that we need to solve only one system of equations (corresponding to the inversion of $M$)
at each time step, which was also already the case for 3D FIT with $\Omega = 0$ (inversion of $M_{\varepsilon e}$).

\section{Numeric Simulation of a Rotating Ring Resonator}

\begin{figure}
\centering
\begin{tikzpicture}[scale=.3]
 \draw (0,0) circle (5);
 \draw (0,0) circle (10);
 \draw (0,0) -- (0,10);
 \node[inner sep=0pt]  at (-.5,3) {$b$};
 \draw (0,0) -- (3.53,3.53);
 \node[inner sep=0pt]  at (2.2,1.5) {$a$};
\end{tikzpicture}
\caption{Geometry of the resonator. $a = 5 \text{mm}$, $b= 10 \text{mm}$.
The height in $z$ dimension is $2 \text{mm}$.}
\label{resonator}
\end{figure}
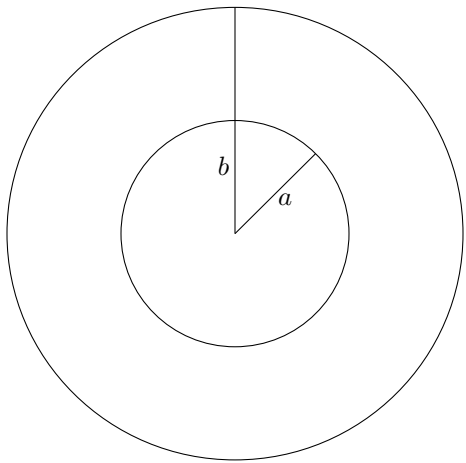

\begin{figure}
    \centering
    \begin{subfigure}[b]{0.3\textwidth}
        \includegraphics[width=\textwidth]{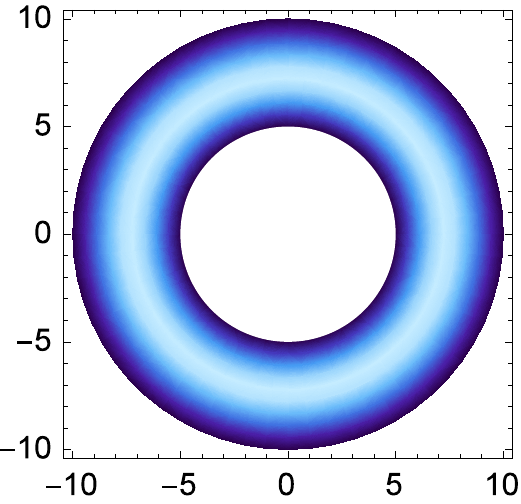}
        \caption{$m=0$}
    \end{subfigure}
    \quad
    \begin{subfigure}[b]{0.3\textwidth}
        \includegraphics[width=\textwidth]{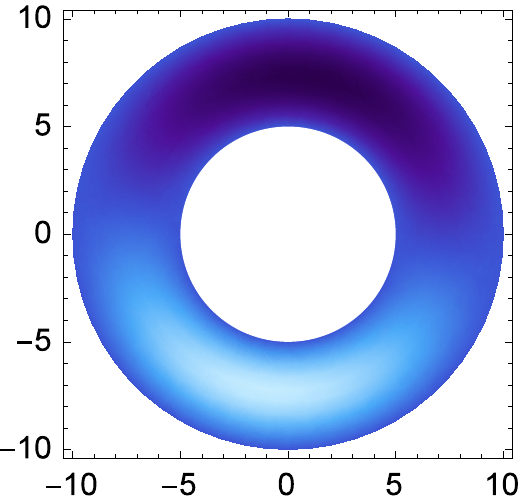}
        \caption{$m=1$}
    \end{subfigure}
    \quad
    \begin{subfigure}[b]{0.3\textwidth}
        \includegraphics[width=\textwidth]{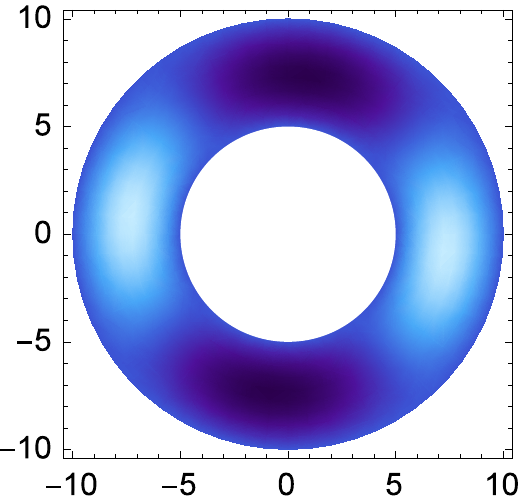}
        \caption{$m=2$}
    \end{subfigure}
    \\
    \begin{subfigure}[b]{0.3\textwidth}
        \includegraphics[width=\textwidth]{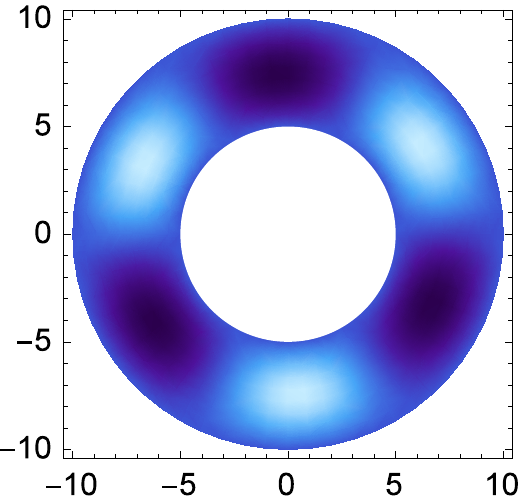}
        \caption{$m=3$}
    \end{subfigure}
    \quad
    \begin{subfigure}[b]{0.3\textwidth}
        \includegraphics[width=\textwidth]{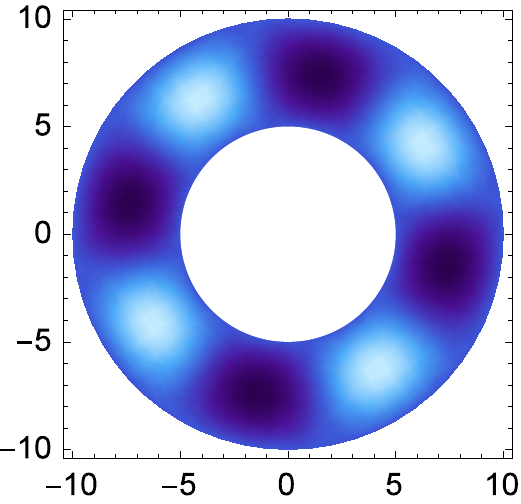}
        \caption{$m=4$}
    \end{subfigure}
    \quad
    \begin{subfigure}[b]{0.3\textwidth}
        \includegraphics[width=\textwidth]{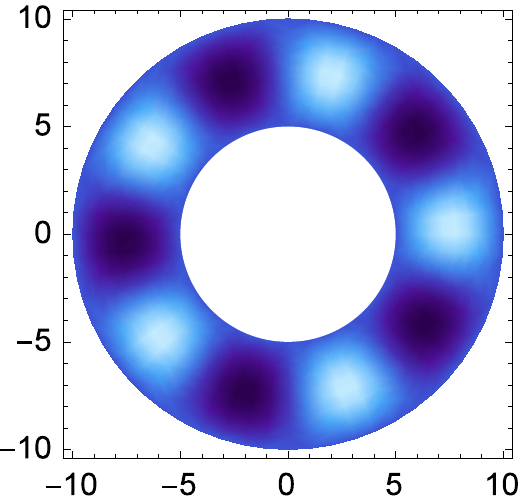}
        \caption{$m=5$}
    \end{subfigure}
    \caption{Considered eigenmodes}
    \label{eigenmodes}
\end{figure}

As a research example we consider the rotating ring resonator depicted in Fig.~\ref{resonator}.
We assume that it is filled with vacuum, and boundary is made of PEC.
First we calculate the eigenmodes of the non-rotating structure using 3D FIT in frequency domain \cite{Weiland1996}.
These are collected in Fig.~\ref{eigenmodes}.
Afterwards they are used as initial values for the time-domain simulation of the rotating structure.

According to the discussion in Section~\ref{sec:gridEquations} the $d$ and $b$ values correspond to
the stationary observer.
Therefore, they can be passed directly (without any processing) from the non-rotating to the rotating solver.
This is
interpreted as exciting the mode field pattern in a stationary frame,
while the PEC boundary is rotating.
One may perceive it that the stationary mode pattern $\vec{E}_\text{s}(x)$ and $\vec{B}_\text{s}(x)$ is transformed 
to the rotating reference frame via local Lorentz transformations, 
yielding $\vec{E}_\text{r}(x) \neq \vec{E}_\text{s}(x)$ and $\vec{B}_\text{r}(x) \neq \vec{B}_\text{s}(x)$.

The other choice would be to insist that the mode pattern should be excited in rotating reference frame, i.e.,
should not be transformed, i.e.,
$\vec{E}_\text{r}(x) = \vec{E}_\text{s}(x)$ and $\vec{B}_\text{r}(x) = \vec{B}_\text{s}(x)$.
This can be achieved by taking
\begin{align}
 e^{1/2}_i: = \int d\tau \bow{e}_i(\tau) \approx \Delta\tau \bow{e}_i(0) = \sqrt{1-v_i^2}\Delta t \bow{e}_i = \frac{\Delta t}{\cosh(r_i\Omega)} \bow{e}_i \\
 h^{0}_i: = \int d\tau \bow{h}_i(\tau) \approx \Delta\tau \bow{h}_i(0) = \sqrt{1-v_i^2}\Delta t \bow{h}_i = \frac{\Delta t}{\cosh(r_i\Omega)} \bow{h}_i \,,
\end{align}
where $v_i$ and thus $r_i$ are values of 3D velocity and radial coordinate at, e.g., the middle of, 
the edge corresponding to $\bow{e}_i$ or $\bow{h}_i$.

Depending on the setting of a particular physical situation one may freely chose
whether the fields should be excited in the rotating or stationary reference frame
by transferring $d$ and $b$ or $e$ and $h$ as initial conditions.

\subsection{Analytic Predictions}

The eigenmodes have harmonic azimuthal and time dependency
\begin{equation}
 E^z_{\text{stat}} = E^z(r,\theta,z,t) = E^z(r,z) \cos(k_\theta\theta)\cos(\omega t) \,.
\end{equation}
They may be perceived as two clockwise and counterclockwise modes
\begin{multline}
 E^z(r,z) \cos(k_\theta\theta)\cos(\omega t) = \\ =
 E^z(r,z) \frac{1}{2}\left[ \cos(k_\theta\theta+\omega t) + \cos(k_\theta\theta-\omega t)  \right] =
 E^+_{\text{stat}} + E^-_{\text{stat}}\,,
\end{multline}
where
\begin{align}
 E^+_{\text{stat}} &:= \frac{1}{2} E^z(r,z) \cos(k_\theta\theta+\omega t) \\
 E^-_{\text{stat}} &:= \frac{1}{2} E^z(r,z) \cos(k_\theta\theta-\omega t) \,.
\end{align}
According to \cite{Steinberg} these two modes should decouple if the structure is rotating
\begin{align}
 E^+_{\text{rot}} &:= 
 \frac{1}{2} E^z(r,z) \cos(k_\theta\theta+\omega_+ t) =
 \frac{1}{2} E^z(r,z) \cos(k_\theta\theta+\omega t + \delta\omega t) 
 \\
 E^-_{\text{rot}} &:= 
 \frac{1}{2} E^z(r,z) \cos(k_\theta\theta-\omega_- t) =
 \frac{1}{2} E^z(r,z) \cos(k_\theta\theta-\omega t + \delta\omega t )
 \,.
\end{align}
Thus imposing initial value $E^z(r,\theta,z,0)$ is equivalent to exciting these modes with the same amplitude.
The solution therefore is
\begin{multline}
 E^z_{\text{rot}}(r,\theta,z,t) = E^+_{\text{rot}} + E^-_{\text{rot}} = \\ =
 \frac{1}{2} E^z(r,z) \left[ 
 \cos(k_\theta\theta+\omega t + \delta\omega t) + 
 \cos(k_\theta\theta-\omega t + \delta\omega t ) 
 \right] = \\ =
 E^z(r,z) \cos(k_\theta\theta + \delta\omega t) \cos(\omega t) \,.
\end{multline}
Therefore, if we chose a reference point $(r,\theta,z) = (r_0,0,z_0)$ and $E^z(r_0,z_0) \neq 0$,
we should observe the time signal 
\begin{equation}
 \frac{E^z_{\text{rot}}(r_0,0,z_0,t)}{E^z(r_0,z_0)} = \cos(\delta\omega t) \cos(\omega t) \,.
 \label{referenceSignal}
\end{equation}
According to \cite[Equation (4.5)]{Steinberg} the rotation induced frequency shift should be
\begin{equation}
 \delta\omega \approx m \Omega \,,
 \label{deltaOmega}
\end{equation}
where $m$ is the number of wavelengths in azimuthal direction.

\subsection{Comparison with Analytic Formula}

\begin{figure}
    \centering
\includegraphics[width= .35\linewidth]{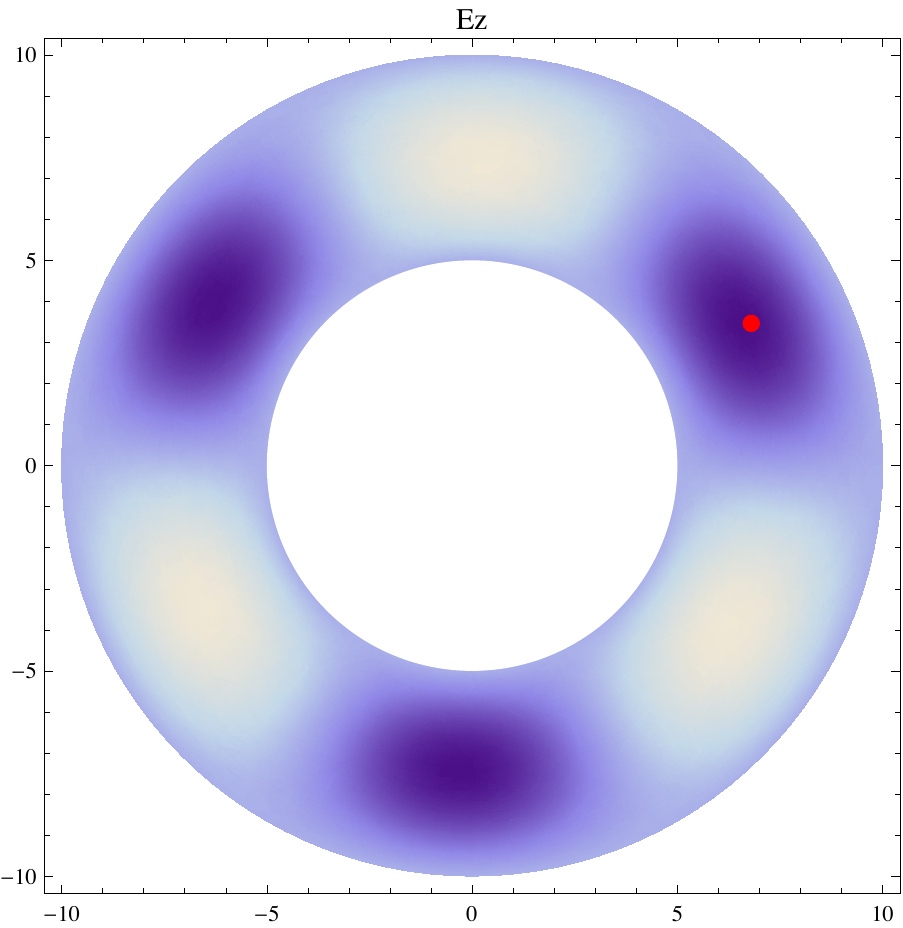}
\hspace{.05\linewidth}
\includegraphics[width= .5\linewidth]{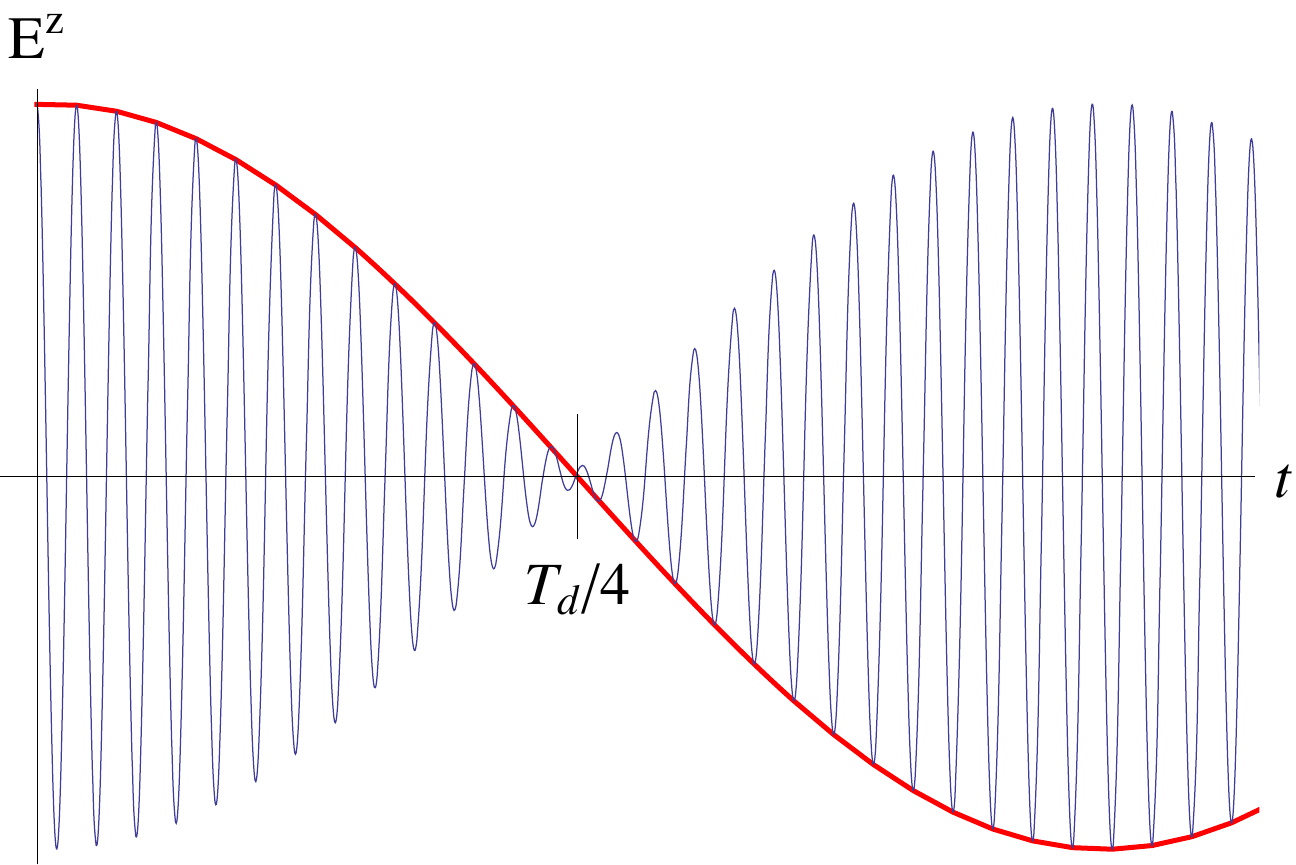}
\caption{Behavior of the electric field
$\vec{E} =\left[0,0,E^z\right] \left[\sigma_x,\sigma_y,\sigma_z\right]^T$ 
at the sample point. $T_d$ is the beating period.}
\label{qualitatively}
\end{figure}

According to \eqref{referenceSignal} the field should look like in Fig.~\ref{qualitatively}.
At time $T_0$ the amplitude is approximately zero and
this point correspond to the quarter of beating period $T_d$.
Therefore, the frequency shift is
\begin{equation}
 \delta\omega = \frac{2\pi}{T_d} = \frac{\pi}{2T_0} \,.
\end{equation}
The frequencies of the discrete time-domain signal are extracted using one of the state of the art algoriths described in
\cite{Pesavento}.

The obtained relative difference with the analytic formula \eqref{deltaOmega}
\begin{equation}
\frac{|\delta \omega^{\text{simulation}} - m \Omega|}{m\Omega}
\label{relativeErr}
\end{equation}
for various $m$ and $\Omega$ is presented in Tables~\ref{tab:FIT} and \ref{tab:FEM}.
\begin{table}
\begin{equation*}
\begin{array}{cc}
  &  v_{\text{max}}/c \\[.5em] m &
\begin{array}{|c|c|c|c|c|c|c|}
\hline & <100 \%&99.63 \%&30.42 \%&3.14 \%&0.31 \%&0.03 \% \\ \hline
0& \text{Null} & \text{Null} & \text{Null} & \text{Null} & \text{Null} & \text{Null} \\ \hline
1& 95.7 \% & 58.1 \% & 2.2 \% & 0. & 0.2 \% & 3.2 \% \\ \hline
2& 95.8 \% & 58.5 \% & 2.8 \% & 0.9 \% & 0.9 \% & 1. \% \\ \hline
3& 95.8 \% & 59.2 \% & 2.6 \% & 2. \% & 3.9 \% & 3.4 \% \\ \hline
4& 96. \% & 60.6 \% & 4.6 \% & 1.3 \% & 1.3 \% & 2.4 \% \\ \hline
5& 96.2 \% & 62.2 \% & 5. \% & 8.2 \% & 8.2 \% & 2.6 \% \\ \hline
\end{array}
\end{array}
\end{equation*}
\caption{FIT case: the relative error \eqref{relativeErr} vs. mode number and rotation rate/velocity of the outer rim of the ring}
\label{tab:FIT}
\end{table}
\begin{table}
\begin{equation*}
\begin{array}{cc}
  &  v_{\text{max}}/c \\[.5em] m &
\begin{array}{|c|c|c|c|c|c|c|}
\hline & <100 \%&99.63 \%&30.42 \%&3.14 \%&0.31 \%&0.03 \% \\ \hline
0& \text{Null} & \text{Null} & \text{Null} & \text{Null} & \text{Null} & \text{Null} \\ \hline
1& 95.6 \% & 57.4 \% & 3.5 \% & 0.4 \% & 6.7 \% & 5.4 \% \\ \hline
2& 95.7 \% & 57.8 \% & 0. & 0.8 \% & 3.5 \% & 3.9 \% \\ \hline
3& 95.7 \% & 58. \% & 0.1 \% & 5.2 \% & 4.9 \% & 5. \% \\ \hline
4& 95.8 \% & 58.4 \% & 0.7 \% & 1.1 \% & 3.4 \% & 7.9 \% \\ \hline
5& 95.8 \% & 58.7 \% & 2.6 \% & 13. \% & 12.9 \% & 2.4 \% \\ \hline
\end{array}
\end{array}
\end{equation*}
\caption{FEM case: the relative error \eqref{relativeErr} vs. mode number and rotation rate/velocity of the outer rim of the ring}
\label{tab:FEM}
\end{table}

\begin{figure}[h]
 \centering
 \includegraphics[width=.6\textwidth]{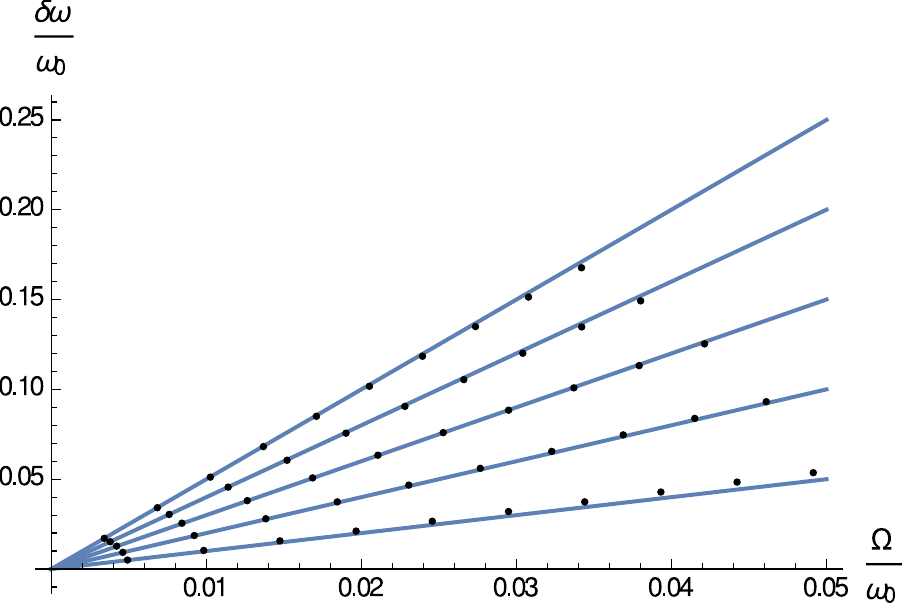}
 \caption{Comparison of analytical and numerical frequency shifts in case of non-relativistic velocities.}
 \label{fig:linearPlot}
\end{figure}

As we can see from Tables~\ref{tab:FIT} and \ref{tab:FEM} 
the numerical results are in close agreement with the analytic formula \eqref{deltaOmega} for velocities $<30\%$ speed of light.
This can also be seen from Fig.~\ref{fig:linearPlot}.
This is expected as this formula was derived assuming that the rotation rate $\Omega$ is low, i.e., $r \Omega/c \ll 1$.
We want to stress that the two solutions: analytic and numerical one suffer from two different types of error.
The analytic solution suffers from the modelling error, namely, absence of relativistic effects.
The numerical solution was obtained without any non-relativistic assumptions.
Therefore, it suffers only from the error coming from the discretization.

\subsection{Extrapolation Leads to Instabilities}

\begin{figure}
    \centering
    \begin{subfigure}[b]{0.7\textwidth}
        \includegraphics[width=\textwidth]{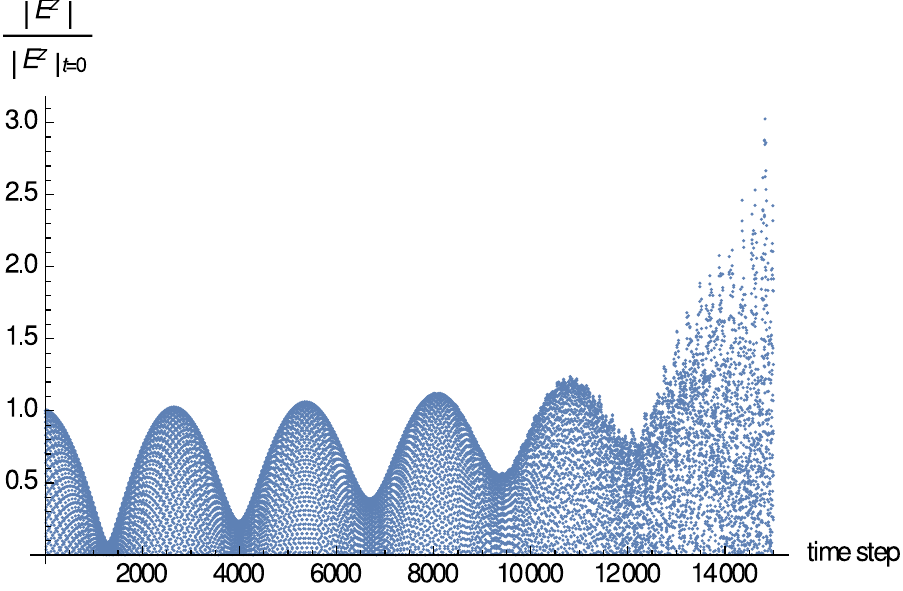}
        \caption{Zeroth order}
    \end{subfigure}
    \\
    \begin{subfigure}[b]{0.7\textwidth}
        \includegraphics[width=\textwidth]{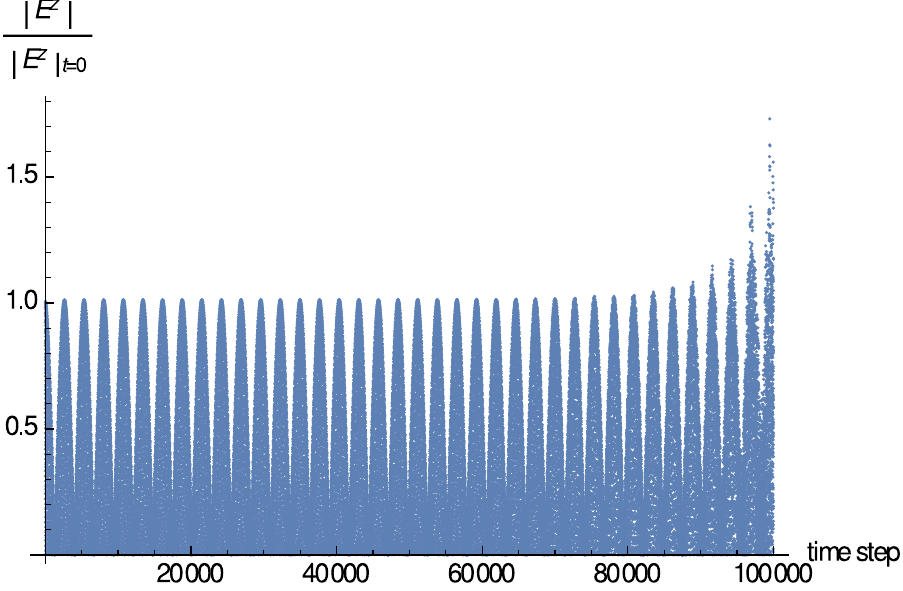}
        \caption{First order}
    \end{subfigure}
    \\
    \begin{subfigure}[b]{0.7\textwidth}
        \includegraphics[width=\textwidth]{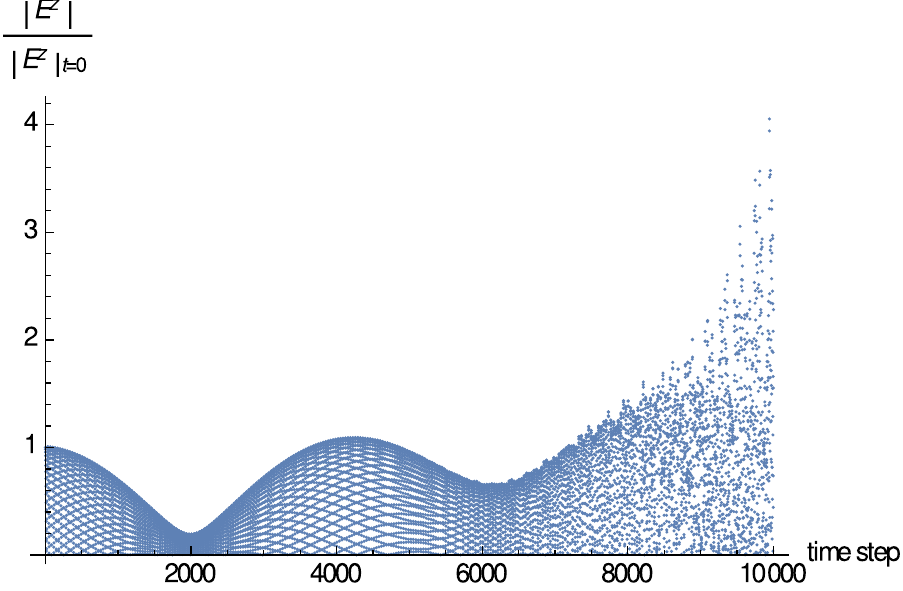}
        \caption{Second order}
    \end{subfigure}
 \caption{Amplitude of the field for various extrapolations (FIT).}
 \label{extrapolationFIT}
\end{figure}

We consider the mode with $m=3$.
The results for zeroth, first and second order extrapolation in our 4D extension of FIT
are depicted in Fig.~\ref{extrapolationFIT}.
We can see that although first order extrapolation is more stable than zeroth one,
increasing the extrapolation order to two gives less stable result.
Therefore, suggestion in \cite{SagnacFDTD} to apply more accurate extrapolation
is not an option as it leads to more unstable scheme.

\begin{figure}
 \centering
 \includegraphics[width=0.7\textwidth]{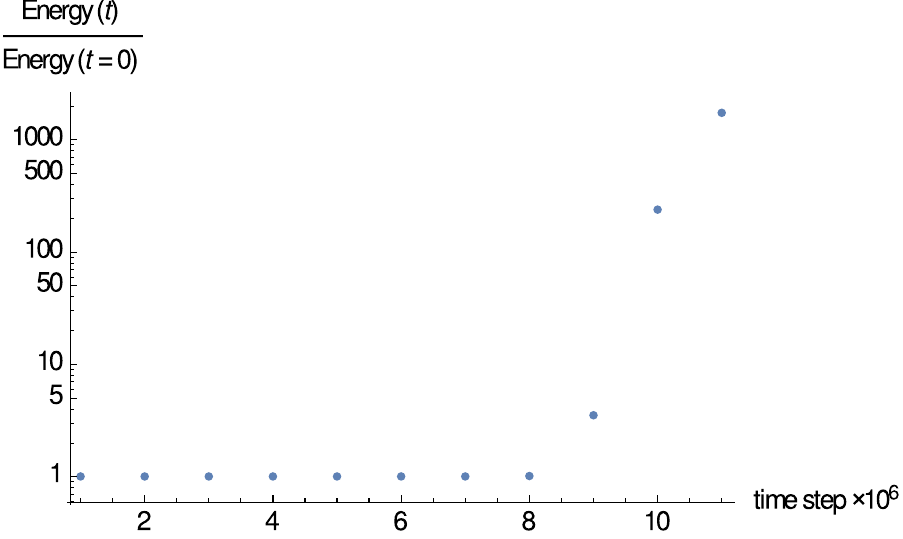}
 \caption{Energy conservation for FIT}
 \label{implicitFIT}
\end{figure}

Applying an implicit scheme with FIT matrices leads to
unstable scheme as we discussed it before.
However, it remains stable for a few millions of time steps as presented in Fig.~\ref{implicitFIT}.
Therefore, extrapolation increases the overall instability of the scheme.

\begin{figure}
    \centering
    \begin{subfigure}[b]{0.7\textwidth}
        \includegraphics[width=\textwidth]{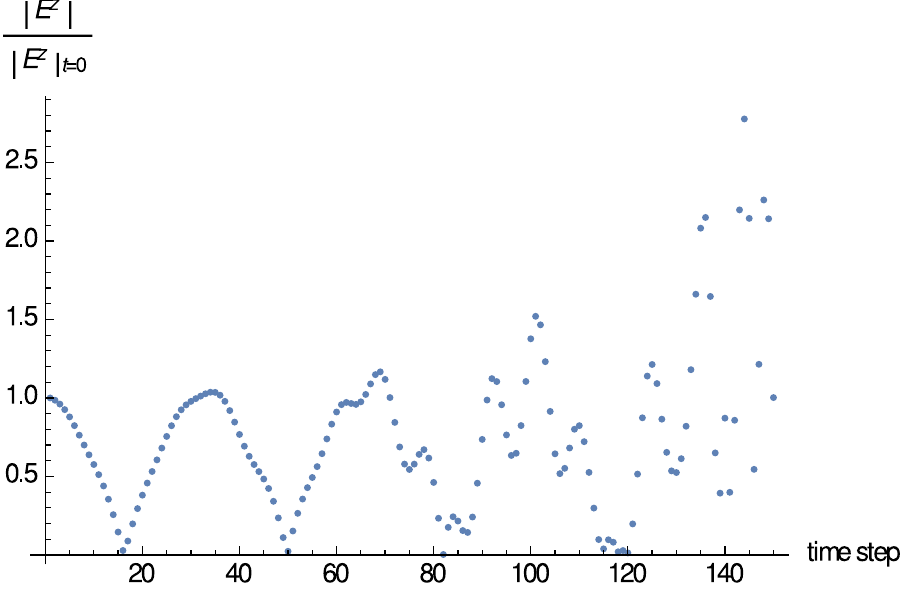}
        \caption{Zeroth order}
    \end{subfigure}
    \\
    \begin{subfigure}[b]{0.7\textwidth}
        \includegraphics[width=\textwidth]{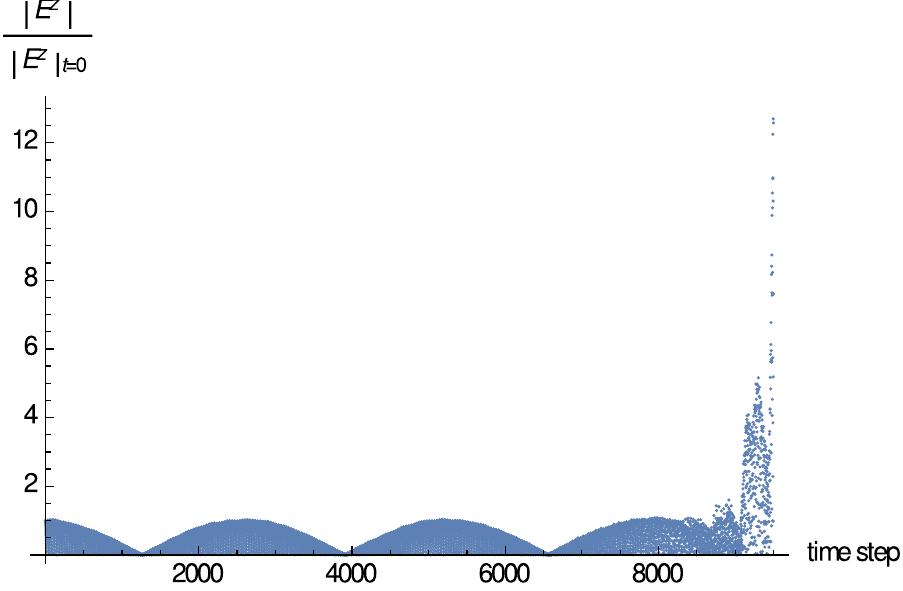}
        \caption{First order}
    \end{subfigure}
    \\
    \begin{subfigure}[b]{0.7\textwidth}
        \includegraphics[width=\textwidth]{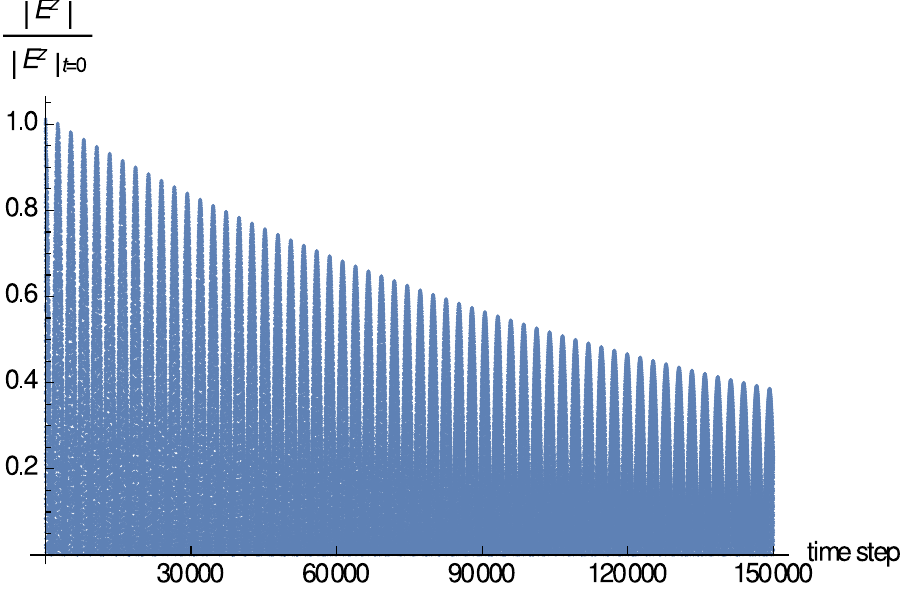}
        \caption{Second order}
    \end{subfigure}
 \caption{Amplitude of the field for various extrapolations (FEM).}
 \label{extrapolationFEM}
\end{figure}

Analogous results for FEM are in Fig.~\ref{extrapolationFEM}.
We see that the zeroth order gets immediately unstable.
First order extrapolation ``explodes'' after 9000 time steps.
Second order extrapolation although stable leads to the dissipation,
which should not be present.
Moreover, there is no guarantee that for other geometries or initial values
this extrapolation will lead to a stable scheme.

\begin{figure}
 \centering
 \includegraphics[width=.7\textwidth]{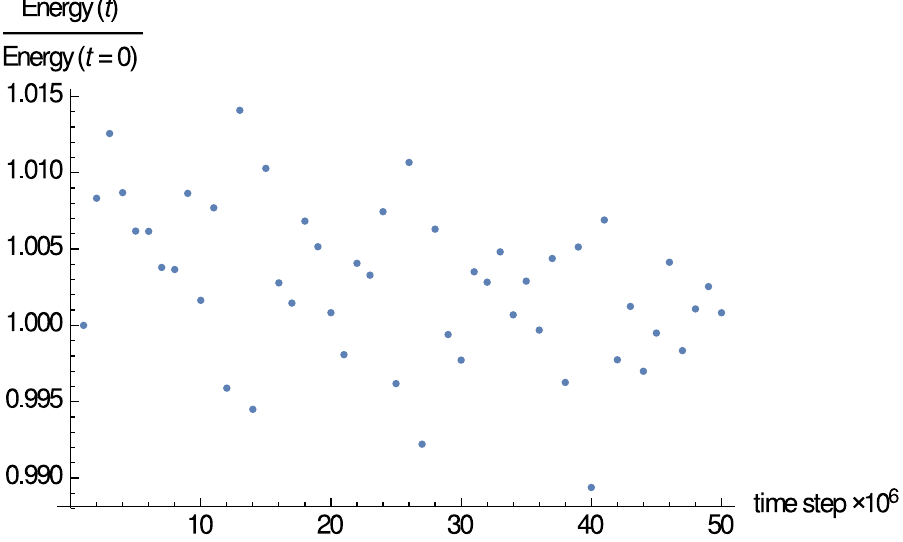}
 \caption{Energy conservation for FEM}
 \label{implicitFEM}
\end{figure}

The implicit FEM scheme remains stable up to 50 millions of time steps
as shown in Fig.~\ref{implicitFEM}.

\section{Conclusion and Outlook}
Starting from the space-time setting we have proposed numerical scheme for simulation of
electromagnetic fields in arbitrary non-inertial reference frame.
We have not made any low-velocity assumptions,
i.e., $\frac{r\Omega}{c},\,\frac{\Omega}{\omega},\,\frac{v}{c} \ll 1$,
at any point, thus
our approach can also be used for a simulation of, e.g.,
linear accelerator or synchrotron radiation.

The application of a space-time method ensures that the obtained scheme is consistent with relativity.
Moreover, it enhances the efficiency of the scheme in the sense that
the material matrices for a rotating observer are constant in time.
The introduced DoFs allow to choose whether the fields should be transformed or not when
passing them from a stationary to a rotating solver.
Our DoFs are also convenient in terms of enforcing boundary conditions.
Not only PEC as discussed in the paper, but also,
e.g., 
one may be interested in the implementation of Absorbing Boundary Conditions (ABCs).
Typically, they are derived by considering outgoing plane waves, which are solutions defined in a stationary reference frame.
Consequently, one would prefer to relate to $d$ and $b$ as they relate to the stationary observer.
For example, extension of Mur's ABCs to a 3D Cartesian grid rotating with arbitrary high rotation rate
is straightforward; namely, one needs to replace $e$ by $d$ in \cite[Eq. 15-16]{Mur}.

Numerical experiments show the necessity of using an implicit scheme (instead of extrapolation of DoFs) in order to
keep stability of the scheme.
Our results both for 4D FEM and FIT are in good agreement with the analytic predictions for low-velocities.

\textbf{Acknowledgement:}
The work of the first, second and third author is supported by the 'Excellence Initiative' 
of the German Federal and State Governments 
and the Graduate School of CE at Technische Universitaet Darmstadt.

\appendix

\section{Spacelike, Lightlike and Timelike}
\begin{equation}
 \mbox{The vector $v$ is called}
 \begin{cases}
 \text{ timelike if } & v > 0   \\
 \text{ lightlike if }& v = 0  \\
 \text{ spacelike if }& v < 0 
 \end{cases}
\end{equation}
This notion naturally extends to $n$-vectors.
We will refer only to 2D facets in this context.
The 2D facet $\Delta$ is called timelike or lightlike if it contains a timelike or lightlike vector, respectively;
otherwise it is called spacelike.
Mathematically, one can test the type of $\Delta$ by squaring the bivector $W$ associated with it; namely
\begin{equation}
 \mbox{The facet $\Delta$ is called}
 \begin{cases}
 \text{ timelike if } & W^2 > 0   \\
 \text{ lightlike if }& W^2 = 0  \\
 \text{ spacelike if }& W^2 < 0 
 \end{cases}
\end{equation}

\section{Velocity Field of a Rotating Observer}
\label{app:velocity}
For simplicity we set $c =1$, obtaining
\begin{align}
 \frac{d p}{d t} &= [1,-\sin(\theta) \tanh(r\Omega),\cos(\theta) \tanh(r\Omega),0][\gamma_t, \gamma_x, \gamma_y, \gamma_z]^T \\
 \left|\frac{d p}{d t}\right| &= \frac{1}{\cosh(r\Omega)} \\
 u &= \frac{d p}{d t} \left/ \left|\frac{d p}{d t}\right| \right. = \nonumber \\ =&
 [\cosh(r\Omega),-\sinh(r\Omega)\sin(\theta),\sinh(r\Omega)\cos(\theta),0][\gamma_t, \gamma_x, \gamma_y, \gamma_z]^T \label{urot} \\
 v_x &= \frac{d x}{d t} = \frac{u_x}{u_t} = -\tanh(r\Omega)\sin(\theta) \\
 v_y &= \frac{d y}{d t} = \frac{u_y}{u_t} = \tanh(r\Omega)\cos(\theta) \\
 v &= \sqrt{v_x^2+v_y^2} = \tanh(r\Omega)
\end{align}

\section{Whitney Interpolation}
Thereby, \eqref{deRhamComplex} requires that
\begin{align}
\nabla\wedge (W^n \underline{A}_n)    &=   W^{n+1} (D^n \underline{A}_n)  \,, \\
\sum\limits_i  \nabla\wedge (A_n^i N^n_i(x))  &=  \sum\limits_{i,\,j} (\left[ D^n\right]_{ji} A_n^i) N^{n+1}_j(x) \,, \\
\sum\limits_i A_n^i \nabla\wedge N^n_i(x)  &=  \sum\limits_{i,\,j} A_n^i \left[ D^n\right]_{ji} N^{n+1}_j(x) \,, \\
\nabla\wedge N^n_i(x)  &=  \sum\limits_j  \left[ D^n\right]_{ji} N^{n+1}_j(x) \,.
\label{WhitneyCondition}
\end{align}

One can indeed verify that Whitney interpolation is compatible with the approximation-free discretization introduced before;
namely, they lead to the same system of equations as we prove next.
In order to obtain the discrete system of equations, we plug
the reconstructed field $F = W^2 \underline{f}$ in $\nabla \wedge F =0$
\begin{equation}
\nabla\wedge F = \sum\limits_{i} f_i \nabla\wedge N^2_i(x) =
\sum\limits_{i,\,j} f_i \left[ D^2 \right]_{ji} N^3_{j}(x) =
\sum\limits_{j} \left( \sum\limits_{i} \left[ D^2 \right]_{ji} f_i \right) N^3_{j}(x) =
0 \,,
\end{equation}
and since the basis functions $N^3_{j}(x)$ are linearly independent, it is equivalent to
\begin{equation}
 \sum\limits_{i} \left[ D^2 \right]_{ji} f_i = 0  \quad \Leftrightarrow \quad \underline{D}^2 \underline{f} = 0 \,,
\end{equation}
which is exactly \eqref{ME4Dgrid}-left.

\section{Explicit Formulae for Whitney Elements}
\label{sec:ExplicitWhitney}
Let us now elaborate on the explicit construction of the basis functions.
Since we consider one 4D element, there is only one basis function associated with a $4$-vector, that is
\begin{equation}
 N^4_1 := \overset{(4)}{I} \,,
\end{equation}
where
\begin{equation}
 \overset{(4)}{I} := \gamma^t \wedge \gamma^x \wedge \gamma^y \wedge \gamma^z \,.
\end{equation}

Similarly, there are eight 3D facets and thus eight basis functions associated with $3$-vectors, viz.
\begin{equation}
 N^3_{i\pm} := \pm \gamma_i \overset{(4)}{I} (1\pm x^i) ,\, i=t,x,y,z \,.
\end{equation}
We can check that \eqref{WhitneyCondition} with $n=3$ holds and 
the rightmost part of the deRham complex \eqref{deRhamComplex} commutes, namely
\begin{equation}
 \nabla\wedge N^3_{i\pm} = 
 \sum\limits_{k}(\gamma^k\partial_k)\wedge \left( \pm \gamma_i \overset{(4)}{I} (1\pm x^i) \right) =
 \sum\limits_{k}\gamma^k\wedge(\gamma_i \overset{(4)}{I}) \delta^i_k  =
 \overset{(4)}{I} = N^4_1 \,.
\end{equation}

There are $24$ 2D facets and thus $24$ basis functions associated with $2$-vectors, viz.
\begin{equation}
 N^2_{i \pm_1,\,j\pm_2} := \frac{1}{16} \gamma_i \wedge \gamma_j \overset{(4)}{I} (1\pm_1x^i)(1\pm_2x^j) \,.
 \label{Whitney2D}
\end{equation}
We can verify with \eqref{WhitneyCondition} and $n=2$ that \eqref{Whitney2D} are indeed Whitney forms
\begin{multline}
 \nabla\wedge N^2_{i \pm_1,\,j\pm_2} =
 \sum\limits_{k} \frac{1}{16} \gamma^k\wedge(\gamma_i \wedge \gamma_j \overset{(4)}{I})
 \left[ \pm_1\delta^i_k(1\pm_2x^j) \pm_2\delta^j_k(1\pm_1x^i) \right] = \\ =
 \frac{1}{16} \left[ \pm_1 \gamma_j \overset{(4)}{I}(1\pm_2x^j) \mp_2 \gamma_i \overset{(4)}{I} (1\pm_1x^i)  \right] =
 \pm_1\pm_2 N^3_{j\pm_2} \mp_2\pm_1 N^3_{i\pm_1} \,.
\end{multline}

In a similar fashion, we can define $N^0$ and $N^1$. 
Nevertheless, since we never refer to them, we do not state them explicitly.

\section{Derivation of $\Phi_i$}
\label{sec:Phi}
The reference tesseract described above has to be mapped to a deformed tesseract in the physical domain.
This is accomplished with the transformation
\begin{align}
  t &= \sum\limits_{i=1}^{16} T_i(\bar{t},\bar{x},\bar{y},\bar{z})  t_i \label{diffeoStart} \\
  x &= \sum\limits_{i=1}^{16} T_i(\bar{t},\bar{x},\bar{y},\bar{z})  x_i \\
  y &= \sum\limits_{i=1}^{16} T_i(\bar{t},\bar{x},\bar{y},\bar{z})  y_i \\
  z &= \sum\limits_{i=1}^{16} T_i(\bar{t},\bar{x},\bar{y},\bar{z})  z_i \,, \label{diffeoEnd}
\end{align}
where $(\bar{t},\bar{x},\bar{y},\bar{z})$ are the coordinates in the reference tesseract,
and $(t,x,y,z)$ those in the physical one, 
and $(t_i,x_i,y_i,z_i)$ are the physical coordinates of the $i$-th vertex.
The functions $T_i$ are given by
\begin{equation}
 T_i(\bar{t},\bar{x},\bar{y},\bar{z}) = \frac{1}{16}(1+\bar{t}_i \bar{t})(1+\bar{x}_i \bar{x})(1+\bar{y}_i \bar{y})(1+\bar{z}_i \bar{z}) \,,
\end{equation}
where $(\bar{t}_i,\bar{x}_i,\bar{y}_i,\bar{z}_i) = (\pm1,\pm1,\pm1,\pm1) $ are the reference coordinates of the $i$-th vertex.

This transformation comes from the following reasoning, compare \cite[Section 8.1.3]{FEM}.
We postulate the map between physical coordinates $\bar{x}^i$ and the reference ones $x^i$
\begin{equation}
 \bar{x}^m = A^m + 
 \sum\limits^4_{i} A^m_i x^i +
 \sum\limits^4_{i<j} A^m_{ij} x^i x^j +
 \sum\limits^4_{i<j<k} A^m_{ijk} x^i x^j x^k +
 \sum\limits^4_{i<j<k<l} A^m_{ijkl} x^i x^j x^k x^l \,.
 \label{diffeoAs}
\end{equation}
There are in total $4 \times (1+4+6+4+1) = 64 $ parameters associated with $A^m,\, A^m_i,\, A^m_{ij},\, A^m_{ijk},\, A^m_{ijkl}$.
Then we impose that this mapping should map the nodes of the reference tesseract to the physical ones.
This accounts for $64$ linear equations as there are $16$ nodes with $4$ coordinates each.
This linear system is composed of $64$ equations with $64$ unknowns, 
and can be solved for $A$'s in terms of positions $\bar{x}^i_j$ of the nodes in the physical tesseract.
By substituting the solution for $A$'s in \eqref{diffeoAs},
we obtain the transformations stated in \eqref{diffeoStart}--\eqref{diffeoEnd}.

For illustration purposes we discuss here the properties of the linear interpolation in 1D.
Coordinate transformation is specified by the function $x(\bar{x})$
\begin{equation}
 x = \sum\limits_{i=1}^2 T_i(\bar{x}) x_i \,,
 \label{mapping1D}
\end{equation}
with 
\begin{equation}
 T_i = \frac{1}{2}(1+\bar{x}_i \bar{x}) \,,
\end{equation}
where $\bar{x}_i=\pm 1$.
Being more explicit, we write
\begin{equation}
 x = x_1 \frac{1}{2}(1-\bar{x}) + x_2 \frac{1}{2}(1+\bar{x}) =
 \frac{1}{2} (x_1 + x_2) + \frac{\bar{x}}{2}(x_2 - x_1 ) \,.
\end{equation}
The mapping \eqref{mapping1D} maps the reference domain $\bar{x} \in [-1,+1]$ to the physical domain
$x \in [x_1,x_2]$. Moreover, the boundary of the reference domain, e.g., the point $\bar{x}=-1$,
is mapped to the boundary of the physical domain, e.g., $x=x_1$.
Also the reference barycenter $\bar{x}=0$ is mapped to the physical barycenter $x=\frac{1}{2} (x_1 + x_2)$.
\\
The linear interpolation is extended to 4D by interpolating four coordinate functions $x^j(\bar{t},\bar{x},\bar{y},\bar{z})$
defined on a tensor product of the 1D reference interval, that is 
$\Xi = [-1,+1] \otimes[-1,+1] \otimes[-1,+1] \otimes[-1,+1]$.
We note that not only the barycenter of the reference tesseract $0\otimes0\otimes0\otimes0$
is mapped to the barycenter of the physical 4D cell, but also
the barycenters of the lower dimensional reference elements, e.g., 
the barycenter $1 \otimes 1 \otimes 0 \otimes 0 $ of the 2D facet $1 \otimes 1 \otimes[-1,+1] \otimes[-1,+1]$,
are mapped to the barycenters of the corresponding physical elements $K_n^i$.

\section{The Transformation to the Physical Domain}
\label{sec:technicalTransformation}
Manipulations below correspond to calculating the pullback of a differential form with respect to a diffeomorphism.
The $N^2_i$ should transform as a $2$-form.
For example in our implementation, for the function
\begin{equation}
 N^2_{x+,y+} = \frac{1}{16} \bar{\gamma}_x \wedge \bar{\gamma}_y I (1+x)(1+y)
\end{equation}
we have used
\begin{equation}
 \bar{\gamma}_x \wedge \bar{\gamma}_y I = 
 \bar{\gamma}_x \cdot (\bar{\gamma}_y \cdot I) = 
 \bar{\gamma}_x \cdot (\bar{\gamma}^t \wedge \bar{\gamma}^x \wedge \bar{\gamma}^z) =
 - \bar{\gamma}^t \wedge \bar{\gamma}^z = (\nabla \bar{z}) \wedge (\nabla \bar{t}) \,.
\end{equation}
We need to express $\nabla \bar{x}^\alpha$ in terms of basis vectors $\gamma^\alpha$ of physical space
\begin{equation}
 \bar{\gamma}^\alpha = \frac{\partial \bar{x}^\alpha}{\partial x^\beta} \gamma^\beta \,.
\end{equation}
However, since we do not have an analytic expression for the inverse mapping $\bar{x}^\alpha (x^\beta)$, 
we need to use
\begin{equation}
 \frac{\partial \bar{x}^\alpha}{\partial x^\beta} =
 \left[ \text{inv}\left( \frac{\partial x}{\partial \bar{x}} \right)\right] ^\alpha_\beta \,,
\end{equation}
where $\text{inv}\left( \frac{\partial x}{\partial \bar{x}} \right)$ is the inverse of the Jacobi matrix with the entries
$\left[\frac{\partial x}{\partial \bar{x}}\right]^\alpha_\beta = \frac{\partial x^\alpha}{\partial \bar{x}^\beta}$.
After making all this substitutions we obtain $N^2_i$ expressed in terms of the basis vectors
associated with Minkowski space-time and its metric, but the reference coordinates are used.

\end{document}